\documentclass[aps, prd, superscriptaddress, eqsecnum, nofootinbib, showkeys, twocolumn, showpacs, preprintnumbers, amsmath, amssymb]{revtex4-1}

\usepackage[usenames, dvipsnames]{xcolor}

\usepackage{graphicx}

\usepackage{gensymb}

\usepackage{tabularx}

\begin{document}
	
\title{What initial condition of inflation would suppress the large-scale CMB spectrum?}

	\author{Pisin Chen}
		\email{pisinchen@phys.ntu.edu.tw}
		\affiliation{Department of Physics, National Taiwan University, Taipei 10617, Taiwan}
		\affiliation{Leung Center for Cosmology and Particle Astrophysics, National Taiwan University, Taipei 10617, Taiwan}
		\affiliation{Kavli Institute for Particle Astrophysics and Cosmology, SLAC National Accelerator Laboratory, Stanford University, Stanford, California 94305, USA}
		\affiliation{Graduate Institute of Astrophysics, National Taiwan University, Taipei 10617, Taiwan}
	
	\author{Yu-Hsiang Lin}
		\email{d00222001@ntu.edu.tw}
		\affiliation{Department of Physics, National Taiwan University, Taipei 10617, Taiwan}
		\affiliation{Leung Center for Cosmology and Particle Astrophysics, National Taiwan University, Taipei 10617, Taiwan}
		\affiliation{Kavli Institute for Particle Astrophysics and Cosmology, SLAC National Accelerator Laboratory, Stanford University, Stanford, California 94305, USA}
	
	\date{\today}


\begin{abstract}

	There is an apparent power deficit relative to the $\Lambda$CDM prediction of the CMB spectrum at large scales, which, though not yet statistically significant, persists from WMAP to Planck data. Proposals that invoke some form of initial condition for the inflation have been made to address this apparent power suppression, albeit with conflicting conclusions. By studying the curvature perturbations of a scalar field in the FLRW universe parameterized by the equation of state parameter $w$, we find that the large-scale spectrum at the end of inflation reflects the super-horizon spectrum of the initial state. The large-scale spectrum is suppressed if the universe begins with the adiabatic vacuum in a super-inflation ($w < -1$) or positive-pressure ($w > 0$) era. In the latter case, there is however no causal mechanism to establish the initial adiabatic vacuum. On the other hand, as long as the universe begins with the adiabatic vacuum in an era with $-1 < w < 0$, even if there exists an intermediate positive-pressure era, the large-scale spectrum would be enhanced rather than suppressed. We further calculate the spectrum of a two-stage inflation model with a two-field potential and show that the result agrees with that obtained from the \emph{ad hoc} single-field analysis.

\end{abstract}

\maketitle


\section{Introduction}
	
	\label{sec:Introduction}
	
	The $\Lambda$CDM model of cosmology with an early inflationary era is very successful in explaining the cosmic microwave background (CMB) power spectrum. However, it has been observed in the \emph{COBE} data that the quadrupole power is lower than the model prediction \cite{Smoot1992, Tegmark1996}. This observation is further confirmed by \emph{WMAP}, reporting the quadrupole power lower than the theoretical expectation by more than $1\sigma$ but less than $2\sigma$ \cite{Bennett2013}. Although this stand-alone low quadrupole mode may be explained by the cosmic variance, the \emph{Planck} observation analyzed the low-$\ell$ ($\ell < 30$) and high-$\ell$ ($\ell \geq 30$) spectra separately, and showed that the best-fit amplitude for the low-$\ell$ spectrum is 10\% lower than that for the high-$\ell$ one at 2.5--3$\sigma$ significance \cite{Planck2013XVa, Planck2015XI}.%
		\footnote{This low-$\ell$/high-$\ell$ tension is present even when the particularly low quadrupole mode is excluded from the analysis \cite{Planck2013XVa}. In \cite{Planck2015XI} it is further pointed out that the low-$\ell$ power deficit is mainly caused by the low multipoles between $\ell = 20$ and 30.}
	
	There have been attempts to explain the low-$\ell$ power suppression of the CMB by introducing some pre-inflation era that breaks the slow-roll condition at about 60 e-folds before the end of inflation \cite{Contaldi2003, Wang2008, Scardigli2011, Bouhmadi-Lopez2013, Kouwn2015, Jain2009, Dudas2012, Namjoo2012, White2014, Biswas2014}. The basic argument about how an era that deviates from the slow-roll dynamics could suppress the power is that the amplitude of the curvature perturbation, $\mathcal{R} \sim H \delta \phi / \dot{\phi}$, would decrease as $|\dot{\phi}|$ increases. This scenario is first realized in the single-field chaotic inflation with potential $V = m^2 \phi^2 / 2$, where $m$ is the mass of the inflaton. If the inflaton $\phi$ starts with large speed $\dot{\phi}^2 \gg m^2 \phi^2$, the kinetic energy dominates the pre-inflation universe, and the power at the horizon scale is suppressed \cite{Contaldi2003}. Other scenarios of violating the slow-roll evolution include the pre-inflation era filled with some radiation \cite{Wang2008}, the primordial black hole remnants \cite{Scardigli2011}, or the frustrated network of topological defects \cite{Bouhmadi-Lopez2013}. There are also pre-inflation models in which the universe is dominated by the spatial curvature as the emergent property from a number of moduli fields in the models of solid inflation \cite{Endlich2013, Kouwn2015}. All of the models above report power suppression at the large scales.
	
	On the other hand, the existence of the pre-inflation decelerating era in models that predict multi-stage inflation does not always result in power suppression \cite{Polarski1992, Ashoorioon2006, Ashoorioon2009, Jain2009, Yamauchi2011, Dudas2012, Namjoo2012, White2014}. In the presence of two fields with mass hierarchy, there are two inflationary eras connected by a decelerating era. With the second inflationary era identified as the last 60 e-folds of the inflation, it is shown that the power is enhanced, rather than suppressed, at large scales that cross the horizon during the first inflationary era or the decelerating era \cite{Polarski1992}. Similar evolution also occurs in the early times of the hybrid inflation \cite{Linde1994}, in which the heavy field is played by the ``waterfall field'' who acquires the mass through the coupling to the inflaton field. In this case, it is however inferred that when the coupling term dominates at the early times, the inflaton field rolls faster due to the coupling, and eventually leads to the power suppression \cite{Contaldi2003}. Among other models of multi-stage inflation which commonly have a decelerating era before the last inflationary era, some predict power suppression at large scales \cite{Jain2009, Dudas2012, White2014}, while some predict enhancement \cite{Polarski1992, Ashoorioon2006, Ashoorioon2009, Yamauchi2011, Namjoo2012}. One therefore naturally wonders:~What initial condition of inflation generated by the pre-inflation era would actually suppress the CMB power spectrum at large scales?
	
	In this work, we address the question through the following steps. We first find the spectrum of the adiabatic vacuum in the universe with a constant equation of state driven by a scalar field. The conditions of having the blue-tilted, red-tilted, or scale-invariant spectra at the super-horizon scales are found. The spectra obtained are based on the assumption that the mode solutions approach the Minkowski limit at small scales. We point out that in the decelerating universe the super-horizon modes would enter the horizon and become sub-horizon, which means that these super-horizon modes are initially causally disconnected. Such assumption in an initially decelerating universe therefore relies on the final state of the mode evolution to govern its initial state, which reverses the cause and effect. Later it was shown that the large-scale power suppression in models with pre-inflation decelerating era is actually a consequence of this unnatural yet widely adopted assumption.%
		\footnote{Such a choice of the initial state for inflation, often referred to as the Bunch-Davies vacuum, even if there exists a non-slow-roll pre-inflation phase, has been commonly assumed in the literature (see, for example, \cite{Contaldi2003, Ashoorioon2006, Wang2008, Namjoo2012, Bouhmadi-Lopez2013, Biswas2014, Kouwn2015}). There also exist numerous proposals of a non-Bunch-Davies vacuum as the basis of the initial condition for a universe that does not begin with a slow-roll phase (see, for example, \cite{Sriramkumar2005, Boyanovsky2006, Holman2008, Agullo2011}).}
	
	In the next step, we demonstrate that for the universe experiencing several eras with different equations of state, the large-scale spectrum is determined by the earliest era in which the universe begins. Starting with a single slow-roll era with the scale-invariant super-horizon spectrum, we find how the spectrum changes as one incrementally stacks a kinetic era, and yet another slow-roll era, into the early times.%
		\footnote{The effect due to piecewise changes of the model parameter was studied in \cite{Mukhanov1991}, where the time-dependent effective inflaton mass was considered.} %
		If the universe begins with the initial adiabatic vacuum in the kinetic era, the spectrum is suppressed at large scales, and we find that the suppression is a direct consequence of the blue-tilted super-horizon spectrum in the initial kinetic era. With another slow-roll era preceding the kinetic era, the large-scale spectrum is enhanced because the super-horizon spectrum in the initial slow-roll era is scale-invariant with the amplitude higher than that generated in the second slow-roll era. In this case, the intermediate kinetic era only serves to connect the two scale-invariant spectra of the two slow-roll eras. One sees that the power suppression stems from the initial blue-tilted super-horizon spectrum, and once the initial spectrum is different, the large-scale power may not be suppressed even if there is a pre-inflation kinetic phase.
	
	We also investigate the scenario that the universe starts with a super-inflation era before the slow-roll inflation. The super-inflation era can be induced in theories of quantum gravity \cite{Tsujikawa2004, Ashtekar2010, Biswas2014, Domenech2015}, or by a scalar field that violates the dominant energy condition \cite{Piao2004a, Baldi2005}. Models in the latter case generally suffer from quantum instabilities and should only be regarded as effective theories (for related discussions, see, for example, \cite{Carroll2003, Cline2004}). The interest here lies in the fact that, opposite to the case of a single pre-inflation kinetic era, it is causal to assume the initial adiabatic vacuum in the pre-inflation super-inflation era. We found that in this case the large-scale power is also suppressed due to the blue-tilted super-horizon initial spectrum in the super-inflation era. Power suppression due to an early super-inflation era has also been inferred in the models of loop quantum gravity \cite{Tsujikawa2004} or bouncing cosmology \cite{Biswas2014}, while in this work a more systematic treatment to the evolution of perturbations is given.
	
	After understanding the character of the spectrum in the multi-stage inflation using the \emph{ad hoc} single-field analysis, we calculate the spectrum of the curvature perturbations in a two-field model with the given potential. We consider the chaotic potential with a coupling term to the second scalar field, which is similar to the effective potential in the early stage of the hybrid inflation. By numerically solving the equations of motion and using the \texttt{CAMB} code \cite{Lewis2000, Lewis2012}, we show that the large-scale spectra of curvature perturbations and CMB are indeed enhanced due to the initial inflationary era.%
		\footnote{As regards the treatment of the two-stage inflation, our approach is closest to that of \cite{Ashoorioon2006, Ashoorioon2009}, in which a more complicated string-motivated two-field model is considered. In \cite{Polarski1992} the two fields have no direct coupling, and certain approximations are used to obtain the analytical solutions in various regimes of the model parameters. In \cite{Jain2009, Dudas2012, White2014, Yamauchi2011}, the single-field models are used. In \cite{Namjoo2012} the system is also modeled by a single fluid.}
	
	The paper is organized as following. The super-horizon spectrum in the universe with constant equation of state is derived in Sec.~\ref{sec:Scaling}. We demonstrate the transformation of the power spectrum in three different background evolutions in Sec.~\ref{sec:SingleField}. The curvature perturbation and CMB spectra of the two-field inflation are found in Sec.~\ref{sec:TwoField}. We conclude in Sec.~\ref{sec:Conclusion}.


\section{Scaling relation}

	\label{sec:Scaling}
	
	To demonstrate the differences between the accelerating eras and the decelerating ones on the mechanism of generating the curvature perturbation spectrum, we analyze the single-field models with \emph{ad hoc} actions in the given background evolutions. In this section, we first find the single-field model that gives the background evolution with a given $w$, and the corresponding general solutions to the curvature perturbations. We then discuss the common assumption on the small-scale behavior of the solution and its causal character. At the end we find the power spectrum in the background with given $w$ and its power-law relation with respect to the wavenumber $k$.

	
	\subsection{Perturbations with constant equation of motion}
	
		\label{subsec:Solution}
		
		We consider a scalar field $\phi$ in an expanding Friedmann-Lema\^{i}tre-Robertson-Walker (FLRW) universe with the action
		\begin{align}
			\label{eq:Action} S = \int d^4 x \sqrt{-g} P(X, \phi),
		\end{align}
		where the kinetic term
		\begin{align}
			X = -\frac{1}{2} \partial_{\mu} \phi \partial^{\mu} \phi.
		\end{align}
		The signature of the metric is $(-, +, +, +)$, and we adopt the Planck units throughout this paper. The energy-momentum tensor can be put into the form of the perfect fluid,
		\begin{align}
			T_{\mu \nu} = P g_{\mu \nu} + (\rho + P) u_{\mu} u_{\nu},
		\end{align}
		by identifying $P$ as the pressure, and the energy density and the velocity as
		\begin{align}
			\rho = &2 X \frac{\partial P}{\partial X} - P, \\
			u_{\mu} = &\frac{\partial_{\mu} \phi}{\sqrt{2 X}}.
		\end{align}
	
	The background evolution with constant equation of motion, $-1 < w \leq 1$, can be modeled by the Lagrangian,
		\begin{align}
			\label{eq:NormalLagrangian} P = X - V(\phi),
		\end{align}
		where, in the homogeneous and isotropic background, $X = \dot{\phi}^2 / 2 \geq 0$, and assuming the potential $V \geq 0$.%
			\footnote{With \eqref{eq:NormalLagrangian}, the evolution of constant $w > -1$ and given initial values $\phi_i$ and $\dot{\phi}_i$ can be realized by the \emph{ad hoc} potential
			\begin{align}
				V(\phi) = \frac{\dot{\phi}_i^2 (1 - w)}{2 (1 + w)} \exp\left[ \sqrt{24 \pi (1 + w)} \; (\phi - \phi_i) \right]. \notag
			\end{align}} %
		The dots denote the time derivatives. If one only requires $\rho = X + V$ to be positive and allows $V$ to be negative, then one can have $w > 1$ when $-X < V < 0$.%
		\footnote{A negative potential with $w > 1$ may be invoked, for example, in the cyclic universe scenario \cite{Steinhardt2002}. It requires special care to treat the perturbations in such models \cite{Battefeld2015}. If we assume the universe is not cyclic, and starts from a big bang followed by a decelerating era (which includes the case $w > 1$), then, as we point out in this work, the initial adiabatic vacuum is acausal. In the cyclic universe scenario, the density perturbations in the post-bounce expanding phase is proposed to be seeded by the perturbations generated in the pre-bounce contracting phase. However, the treatment of perturbations in the contracting phase and across the bounce is still under investigation. See, for example, \cite{Battefeld2015} for a review.}
	
	In the conformal Newtonian gauge, the perturbed FLRW metric is
		\begin{align}
			ds^2 = -( 1 + 2 \Phi ) dt^2 + a^2(t) ( 1 + 2 \Psi ) d\mathbf{x}^2,
		\end{align}
		where $a$ is the scale factor, and $\Phi$ and $\Psi$ are the metric perturbations. The equation of motion of the curvature perturbation, $\mathcal{R}$, is given by the Mukhanov-Sasaki equation in the Fourier space \cite{Mukhanov1985, Sasaki1986},
		\begin{align}
			\label{eq:MSEq} \mathcal{R}'' + 2 \frac{A'}{A} \mathcal{R}' + k^2 \mathcal{R} = 0,
		\end{align}
		where
		\begin{align}
			\mathcal{R} = &\Psi - \frac{H}{\dot{\phi}} \delta \phi, \\
			A = &\frac{a \sqrt{\rho + P}}{H},
		\end{align}
		$k$ is the wavenumber, $H = \dot{a} / a$ is the Hubble parameter, $\delta \phi$ is the field perturbation. The primes denote the derivative with respect to the conformal time $\eta$, defined by $dt = a d\eta$. Introducing the new variable $u = -A \mathcal{R}$, we can get rid of the first-derivative term, turning \eqref{eq:MSEq} into
		\begin{align}
			\label{eq:MukhanovEq} u'' + \left( k^2 - \frac{A''}{A} \right) u = 0.
		\end{align}
	
	If the evolution of the universe is described by a constant $w > -1$, there is a simple relation $A'' / A = a'' / a$ since
		\begin{align}
			\label{eq:ARelation} A = \sqrt{\frac{3 ( 1 + w )}{8 \pi}} \; a.
		\end{align}
		The scale factor evolves as
		\begin{align}
			\label{eq:ScaleFactor} a(\eta) = a_i ( 1 + \alpha \xi )^{1 / \alpha},
		\end{align}
		where $\xi = a_i H_i ( \eta - \eta_i )$, $a_i$ and $H_i$ denote the initial values at $\eta = \eta_i$, and
		\begin{align}
			\label{eq:alpha} \alpha = \frac{1 + 3 w}{2}.
		\end{align}
		Equation \eqref{eq:MukhanovEq} then reads
		\begin{align}
			\label{eq:MukhanovEq2} \frac{d^2 u}{d \xi^2} + \left[ \kappa^2 - \frac{\beta}{( 1 + \alpha \xi )^2} \right] u = 0,
		\end{align}
		with $\kappa = k / a_i H_i$ and $\beta = ( 1 - 3 w ) / 2$. The general solution is
		\begin{align}
			\label{eq:WhittakerSolution} u(\xi) = C_1 M_{0, \mu} \left[ 2 i \kappa \left( \xi + \frac{1}{\alpha} \right) \right] + C_2 W_{0, \mu} \left[ 2 i \kappa \left( \xi + \frac{1}{\alpha} \right) \right],
		\end{align}
		in which $M_{\nu, \mu}(z)$ and $W_{\nu, \mu}(z)$ are the Whittaker functions, and
		\begin{align}
			\label{eq:mu} \mu = \frac{3}{2} \left| \frac{1 - w}{1 + 3 w} \right|.
		\end{align}
		Note that $w = -1/3$ is a singular case.%
		\footnote{The super-horizon spectrum is asymptotically divergent for $w = -1/3$ (or $\mu = 0$). To understand why, first note that in this case the universe does not accelerate nor decelerate ($\ddot{a} = 0$), so the comoving scale of Hubble horizon is constant in time. If $w$ is slightly smaller than $-1/3$, the universe accelerates but slowly. It takes a long time for the horizon to shrink a little. At the meantime the amplitudes of the fluctuations inside the horizon keep decaying, therefore the amplitude of the power spectrum changes much within a small range of $k$.} %
		Corresponding $\alpha$ and $\mu$ for some reference values of $w$ are listed in Table \ref{tab:EquationOfState}.


\begin{table}
	\caption{\label{tab:EquationOfState} Corresponding values of $\alpha$ and $\mu$ for some reference equation-of-state parameter $w$. The parameter $\alpha = (1 + 3 w) / 2$ is related to the scale factor by \eqref{eq:ScaleFactor}, and $\mu = | 3 ( 1 - w ) / 2 ( 1 + 3 w ) |$ describes the general solution of the perturbation through \eqref{eq:WhittakerSolution}. Note that for the accelerating universe with $w < -1/3$, one has $\alpha < 0$, while for the decelerating universe with $w > -1/3$, one has $\alpha > 0$.}
	\begin{ruledtabular}
	\begin{tabular}{cccccccccc}
	$w$			&$-\infty$			&$-1$			&$\displaystyle{-\frac{2}{3}}$	&$\displaystyle{-\frac{1}{3}}$	&$0$			&$\displaystyle{\frac{1}{3}}$	&$\displaystyle{\frac{2}{3}}$	&$1$	&$+\infty$	\\ \\
	
	$\alpha$	&$-\infty$			&$-1$			&$\displaystyle{-\frac{1}{2}}$	&$0$			&$\displaystyle{\frac{1}{2}}$	&$1$				&$\displaystyle{\frac{3}{2}}$	&$2$	&$+\infty$	\\ \\
	
	$\mu$		&$\displaystyle{\frac{1}{2}}$		&$\displaystyle{\frac{3}{2}}$	&$\displaystyle{\frac{5}{2}}$	&$+\infty$	&$\displaystyle{\frac{3}{2}}$	&$\displaystyle{\frac{1}{2}}$	&$\displaystyle{\frac{1}{6}}$	&$0$	&$\displaystyle{\frac{1}{2}}$	\\
	\end{tabular}
	\end{ruledtabular}
\end{table}


		If we try to model the slow-roll evolution by assigning $w = -1$, we will end up with $A = 0$ and cannot proceed in the way we did in the previous paragraph. The way around that is to use the attractor solution of the slow-roll era. By writing the density and pressure in terms of field, we have
		\begin{align}
			A = -\frac{\phi'}{H},
		\end{align}
		assuming $\phi' < 0$ without loss of generality. In the attractor regime, $H$ as well as $\dot{\phi} = \phi' / a$ are approximately constant, so we can write
		\begin{align}
			\label{eq:SlowRollA} A = -\frac{\dot{\phi}_i}{H} a,
		\end{align}
		which is proportional to $a$ as it is in \eqref{eq:ARelation}. Also it can be verified by solving the Friedmann equation with constant $H$ that \eqref{eq:ScaleFactor} reproduces the scale factor in the slow-roll case, so the equation of motion \eqref{eq:MukhanovEq2} still holds. We will refer to the slow-roll limit as $w \simeq -1$ in this paper.
	
	For the super-inflationary universe with $w < -1$, we model it by the Lagrangian,
		\begin{align}
			\label{eq:PhantomLagrangian} P = -X - V(\phi),
		\end{align}
		with the sign of the kinetic term reversed. With the requirement $\rho = -X + V > 0$, one generally has $V > X > 0$, and the scale factor still evolves as \eqref{eq:ScaleFactor}. By substituting the original definition of $A$ with
		\begin{align}
			A = \frac{a \sqrt{-\rho - P}}{H},
		\end{align}
		it turns out that the equation of motion of the curvature perturbation can still be written in the form of the Mukhanov-Sasaki equation \eqref{eq:MSEq}, and the rest of the analysis follows.

	
	\subsection{Assumption and character of the small-scale solution}
	
		\label{subsec:InitialCondition}
		
		The common assumption on the initial condition is that the mode solution approaches the Minkowski limit in the short-wavelength limit,
		\begin{align}
			\label{eq:MinkowskiLimit} u(\eta) = \frac{1}{(2\pi)^{3/2} \sqrt{2 k}} e^{-i k \eta} \quad \textrm{(for $k \eta \rightarrow \infty$)}.
		\end{align}
		The Whittaker function that matches this form when $z \gg 1$ is
		\begin{align}
			W_{0, \mu}(2 i z) = e^{- i z} \left[ 1 + \mathcal{O}\left( \frac{1}{z} \right) \right].
		\end{align}
		Therefore, the requirement of matching the adiabatic vacuum picks out the solution,
		\begin{align}
			\label{eq:InitialCondition} u(\xi) = \frac{e^{i \kappa / \alpha}}{(2\pi)^{3/2} \sqrt{2 a_i H_i \kappa}} W_{0, \mu}\left[ 2 i \kappa \left( \xi + \frac{1}{\alpha} \right) \right].
		\end{align}
	
	If at the early times of the inflationary history, the universe is initially accelerating and evolves with a constant equation of state, then by assuming the adiabatic vacuum at the sub-horizon limit, we obtain the initial condition \eqref{eq:InitialCondition}. This sub-horizon initial condition, combined with the quasi-de Sitter expansion, predicts the observed nearly scale invariant super-horizon spectrum in the $\Lambda$CDM universe. One of the reasons that make this scenario attractive is that the quantum fluctuations we learn well in the local Minkowski spacetime, after being stretched to the cosmic scale by inflation, also form the seed of the cosmic structure.
	
	When applying the limit \eqref{eq:MinkowskiLimit} to a decelerating universe, this sub-horizon assumption becomes acausal. In the decelerating universe, the Hubble horizon grows faster than the perturbations do, just like in the late-time universe dominated by matter or radiation. The sub-horizon spectrum is therefore formed after the super-horizon modes enter the horizon. The estimation we make to the super-horizon spectrum is then based on that, after the modes enter the horizon, they must fall in the vacuum state at the small-scale limit. Through the analysis in the next section, we will see that the large-scale power suppression caused by the pre-inflation kinetic era actually originates from the super-horizon spectrum deduced from this picture.

	
	\subsection{Power spectrum}
	
		\label{subsec:Power}
		
		The power spectrum of $\mathcal{R}$ is defined through the expectation value of $\hat{\mathcal{R}}^2 ( \bold{x}, t )$,
		\begin{align}
			\langle \hat{\mathcal{R}}^2 ( \bold{x}, t ) \rangle = \int \frac{d k}{k} P(k).
		\end{align}
		Expanding $\mathcal{R}$ in terms of the creation and annihilation operators,
			\begin{align}
				\mathcal{R}(\mathbf{x}, t) = \int d^3 \mathbf{k} \left[ a_{\mathbf{k}} \mathcal{R}_k(t) e^{i \mathbf{k}\cdot\mathbf{x}} + a_{\mathbf{k}}^{\dagger} \mathcal{R}_k^*(t) e^{-i \mathbf{k}\cdot\mathbf{x}} \right],
			\end{align}
			and using the commutator
			\begin{align}
				[ a_{\mathbf{k}}, a_{\mathbf{k}'}^{\dagger} ] = \delta^3(\mathbf{k} - \mathbf{k}'),
			\end{align}
			one finds that
		\begin{align}
			P(k) = 4 \pi k^3 |\mathcal{R}_k|^2.
		\end{align}
		Recalling that $\mathcal{R} = - u / A$, we find that, for $w \neq -1$ (so $\alpha \neq -1$), the power spectrum given by the solution \eqref{eq:InitialCondition} is
		\begin{align}
			\label{eq:power2} P = \frac{H_i^2 \kappa^2 | 1 + \alpha \xi |^{-2 / \alpha}}{\pi | 1 + \alpha |} \left| W_{0, \mu}\left[ 2 i \kappa \left( \xi + \frac{1}{\alpha} \right) \right] \right|^2.
		\end{align}
		For the slow-roll case, we approximate $A$ by the slow-roll limit \eqref{eq:SlowRollA}, and the mode function in that limit is then given by \eqref{eq:InitialCondition} with $w \simeq -1$. In terms of the slow-roll parameter,
		\begin{align}
			\epsilon \equiv &\frac{1}{16 \pi} \left( \frac{1}{V} \frac{\partial V}{\partial \phi} \right)^2 = \frac{4 \pi \dot{\phi}^2}{H^2},
		\end{align}
		the power spectrum at the slow-roll limit is given by
		\begin{align}
			\label{eq:power3} P = \frac{H_i^2 \kappa^2 | 1 + \alpha \xi |^{-2 / \alpha}}{\pi \epsilon} \left| W_{0, \frac{3}{2}}\left[ 2 i \kappa \left( \xi + \frac{1}{\alpha} \right) \right] \right|^2.
		\end{align}
	
	Using the small-argument expansion of the Whittaker function \cite{Olver2010}, the super-horizon limits of the power spectrum are found for different ranges of $w$. For $w < -1/3$ and $w > 1$ except $w = -1$, one has
		\begin{align}
			P_{\kappa \ll 1} = \frac{H_i^2 \Gamma^2( 2 \mu )}{\pi \Gamma^2( \mu + \frac{1}{2} ) | 1 + \alpha |} \left| \frac{\alpha}{2} \right|^{2 \mu - 1} \kappa^{-2 \mu + 3}.
		\end{align}
		For $w \simeq -1$, the slow-roll power spectrum \eqref{eq:power3} recovers the familiar scale-invariant spectrum
		\begin{align}
			P_{\kappa \ll 1} = \frac{H_i^2}{\pi \epsilon}.
		\end{align}
		For $-1/3 < w < 1$, the super-horizon power spectrum decays with time, given by
		\begin{align}
			P_{\kappa \ll 1} = \frac{H_i^2 \Gamma^2( 2 \mu )}{\pi \Gamma^2( \mu + \frac{1}{2} ) | 1 + \alpha |} \left| \frac{\alpha}{2} \right|^{2 \mu - 1} \kappa^{-2 \mu + 3} T(\xi),
		\end{align}
		with the time-dependence
		\begin{align}
			T(\xi) = | 1 + \alpha \xi |^{-6 ( 1 - w ) / ( 1 + 3 w )}.
		\end{align}
		For $w = 1$, the spectrum is
		\begin{align}
			P_{\kappa \ll 1} = \frac{H_i^2}{3 \pi^2} \kappa^3 T(\xi),
		\end{align}
		with
		\begin{align}
			T(\xi) = \left| \ln \left[ 2 i \kappa \left( \xi + \frac{1}{\alpha} \right) \right] + \gamma - 2 \ln 2 \right|^2,
		\end{align}
		where $\gamma$ is the Euler-Mascheroni constant.
	
	We can summarize the power-law relations of the super-horizon power spectrum with respect to the normalized wavenumber $\kappa$ by the scaling relation%
		\footnote{This relation is also derived in \cite{Cai2015} as the approximation to the super-horizon spectrum at the end of the multi-stage inflationary evolution. The authors focus on the recursive matrix formalism of the multi-stage pre-inflationary era, with the assumption that every pre-inflation era is an accelerating expansion (or decelerating contraction in the bounce inflation scenario).}
		\begin{align}
			\label{eq:ScalingRelation} P \propto \kappa^{-2 \mu + 3},
		\end{align}
		where the correspondence between the case of $w \simeq -1$ and the slow-roll limit is understood. For the convenience of reading, we also state this result in terms of the more familiar parameter, $w$. For $w < -1/3$ and $w > 1$,
		\begin{align}
			P \propto \kappa^{6 ( 1 + w ) / ( 1 + 3 w)}.
		\end{align}
		For $-1/3 < w < 1$,
		\begin{align}
			P \propto \kappa^{12 w / ( 1 + 3 w  )}.
		\end{align}
	
\begin{figure}

\centering

\includegraphics[width = 8 cm]{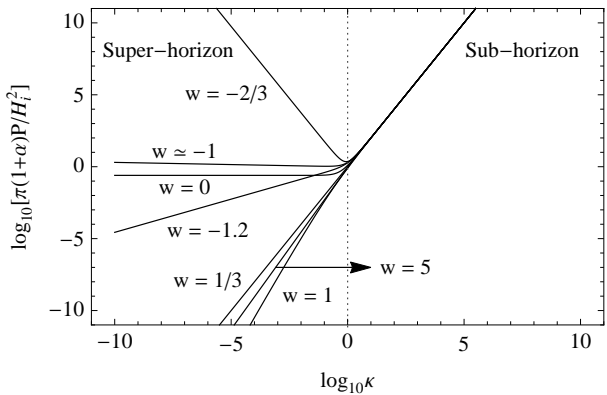}

\caption{The power spectra \eqref{eq:power2} with sample parameters $w = -1.2$, $-2/3$, 0, $1/3$, 1, 5, and at the slow-roll limit $w \simeq -1$. They are plotted with normalizations such that they have the same magnitude at the sub-horizon limit. For $w \simeq -1$ the power spectrum is given by \eqref{eq:power3}, and the normalization is instead $\pi \epsilon P / H_i^2$.}

\label{fig:InitialPower41}

\end{figure}

\begin{figure}

\centering

\includegraphics[width = 8 cm]{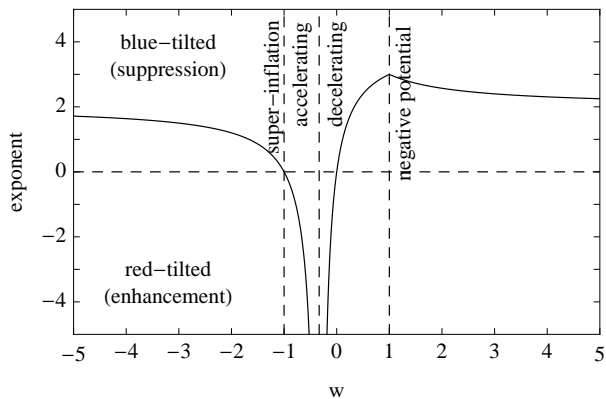}

\caption{The plot of $- 2 \mu + 3$, the exponent of the power-law scaling relation \eqref{eq:ScalingRelation} with respect to the equation-of-state parameter, $w$.}

\label{fig:InitialPower50}

\end{figure}

	The super-horizon behavior of the spectrum can be divided into three types according to the scaling relation. Some typical cases are plotted in FIG.~\ref{fig:InitialPower41}. For $\mu = 3/2$, the spectrum is scale-invariant. This is the case for $w \simeq -1$ (slow-roll) and $w = 0$. When $\mu < 3/2$, the spectrum is blue-tilted and the power is lower than the scale-invariant spectrum at super-horizon scales. This is attainable from the positive-pressure ($w > 0$) or super-inflation era ($w < -1$). In the third case, the super-horizon spectrum is red-tilted, which is achieved when $\mu > 3/2$, or equivalently an era with $-1 < w < 0$ except $w = -1/3$ (FIG.~\ref{fig:InitialPower50}). Here we reiterate that the spectra obtained are based on the adiabatic vacuum \eqref{eq:MinkowskiLimit}, where the corresponding initial condition for the decelerating universe ($w > -1/3$) is acausal.


\section{Evolution of the power spectrum}

	\label{sec:SingleField}
	
	The character of the large-scale spectrum at the end of inflation reflects the nature of the initial state. If at the early times before the onset of inflation, the universe is initially described by some adiabatic vacuum in the background with constant equation of state, the power spectrum is given by \eqref{eq:power2} during that era. As the universe evolves with time, the power spectrum continuously transforms accordingly. The shape of the spectrum at the long-wavelength limit is nevertheless not affected by the evolution, and is preserved in the late-time spectrum.
	
	We demonstrate in this section that, as a consequence of the spectrum evolution, the spectra can be radically different at large scales for distinct initial vacua. Particularly, if the universe transits from a kinetic era into the inflation era, the large-scale spectrum is suppressed because of the initial adiabatic vacuum assumed in the kinetic era. If the initial vacuum is different---for example, changed by an earlier accelerating era before the kinetic era, as discussed in this section---the large-scale spectrum may even become enhanced.
	
	We compare the evolution of the power spectrum in three cases, modeled phenomenologically by the single field dynamics. The first one is a single slow-roll era (denoted as era C) with Hubble parameter $H_C$. The second one is the slow-roll era (era C) preceded by a kinetic era (era B). In the third case we add one more slow-roll era (era A), with Hubble parameter $H_A$, before the kinetic era (era B), which is again followed by the slow-roll era (era C). When applicable, the quantities at the transition from era A to B are denoted by subscript 1 (so, for example, the scale factor at the transition is equal to $a_1$), and those at the transition from B to C are by subscript 2.
	
	In view of the acausal character of the adiabatic vacuum in the initially decelerating era (the second case with only era B and C), we also analyze the case of having a super-inflation era (era S) before the slow-roll era (era C), which also implies power suppression at large scales but is free from the acausal property. Analogously, the quantities at the transition from era S to C are denoted by subscript 2.

	
	\subsection{Slow-roll}
	
	\label{subsec:C}
	
	In the slow-roll era (era C), the general solution to the mode function at the slow-roll limit $w \simeq -1$ is
		\begin{align}
			\label{eq:SolutionC} u_C = \; &C_{+} \left( 1 - i \frac{\tilde{a}_C}{\tilde{k}_C} \right) e^{-i \tilde{k}_C / \tilde{a}_C} \notag \\
			&+ C_{-} \left( 1 + i \frac{\tilde{a}_C}{\tilde{k}_C} \right) e^{i \tilde{k}_C / \tilde{a}_C},
		\end{align}
		where $\tilde{k}_C = k / a_2 H_2$ and $\tilde{a}_C = a / a_2 = [ 1 - a_2 H_2 (\eta - \eta_2) ]^{-1}$. Here $a_2$ and $H_2$ can be viewed as quantities at some reference time, $\eta_2$. The notations are chosen for the convenience of later comparison and should not cause confusion. Note that in the slow-roll era, one can approximate $H_C \approx H_2$ as a constant. The normalized power spectrum is given by
		\begin{align}
			\label{eq:PowerC} &\tilde{P}_C = \frac{\tilde{k}_C^3}{\tilde{a}_C^2} \left| \tilde{C}_{+} \left( 1 - i \frac{\tilde{a}_C}{\tilde{k}_C} \right) e^{-i \tilde{k}_C / \tilde{a}_C} \right. \notag \\
			&\quad\quad\quad\quad\quad\quad\quad \left. + \tilde{C}_{-} \left( 1 + i \frac{\tilde{a}_C}{\tilde{k}_C} \right) e^{i \tilde{k}_C / \tilde{a}_C} \right|^2,
		\end{align}
		where $\tilde{C}_{\pm} = \sqrt{a_2 H_2} C_{\pm}$, and
		\begin{align}
			\tilde{P}_C = &\frac{\epsilon P}{16 \pi^2 H_C^2}, \\
			\epsilon = &\left. \frac{4 \pi \dot{\phi}^2}{H^2} \right|_C.
		\end{align}
		are the normalized power spectrum and the slow-roll parameter evaluated in era C, respectively.
	
\begin{figure}

\centering

\includegraphics[width = 8 cm]{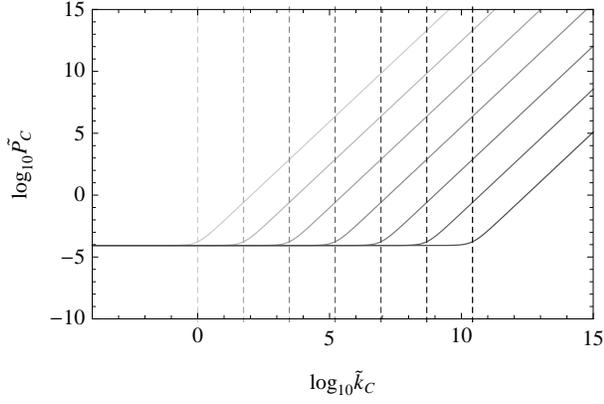}

\caption{Time evolution of the power spectrum in era C in the case of single-stage evolution (only era C). The solid curves are the spectra of $\mathcal{R}$ from early time to late time (from light to dark). The vertical dashed lines denote the comoving horizon size at the corresponding instants (also from light to dark).}

\label{fig:CinC70}

\end{figure}

	In the adiabatic vacuum \eqref{eq:MinkowskiLimit}, only the second term in \eqref{eq:SolutionC} remains, and the mode function reduces to
		\begin{align}
			u_C = \frac{1}{(2 \pi)^{3/2} \sqrt{2 k}} \left( 1 + i \frac{\tilde{a}_C}{\tilde{k}_C} \right) e^{i \tilde{k}_C / \tilde{a}_C}.
		\end{align}
		The power spectrum is
		\begin{align}
			\tilde{P}_C = \frac{1}{16 \pi^3} \left( 1 + \frac{\tilde{k}_C^2}{\tilde{a}_C^2} \right),
		\end{align}
		which recovers the well-known form in the super-horizon limit,
		\begin{align}
			\label{eq:SlowRollSpectrum} P = \frac{1}{\pi} \left( \frac{H_C^2}{\epsilon} \right). \quad \textrm{( $\tilde{k}_C \ll 1$ at $\tilde{a}_C = 1$ )}
		\end{align}
		The comoving horizon size decreases with time, moving toward the right to the small scales in FIG.~\ref{fig:CinC70}. After the mode exits the horizon, lying on the left-hand side of the horizon scale, the power stays scale-invariant.


	\subsection{Kinetic---slow-roll}
	
	\label{subsec:BC}
	
		In the second case, the energy density is dominated by the kinetic energy at $\eta < \eta_2$, with
		\begin{align}
			H \approx -\sqrt{ \frac{4 \pi}{3} } \frac{\phi'}{a},
		\end{align}
		assuming $\phi' \leq 0$ without lose of generality. In the kinetic era, $w = 1$ and the mode function can be written in terms of the Hankel functions,
		\begin{align}
			\label{eq:SolutionB} u_B = &B_{+} \tilde{a}_B H_0^{(1)}\left( \frac{1}{2} \tilde{k}_B \tilde{a}_B^2 \right) \notag \\
			&+ B_{-} \tilde{a}_B H_0^{(2)}\left( \frac{1}{2} \tilde{k}_B \tilde{a}_B^2 \right),
		\end{align}
		where we denote $\tilde{k}_B = k / a_1 H_1$ and $\tilde{a}_B = a / a_1 = \sqrt{1 + 2 a_1 H_1 ( \eta - \eta_1 )}$. Similarly, for later convenience, we choose $\eta_1 < \eta_2$ to be some reference time for era B. The power spectrum in the kinetic era is given by
		\begin{align}
			\label{eq:PowerB} &\tilde{P}_B = \tilde{k}_B^3 \left| \tilde{B}_{+} H_0^{(1)}\left( \frac{1}{2} \tilde{k}_B \tilde{a}_B^2 \right) \right. \notag \\
			&\quad\quad\quad\quad \left. + \tilde{B}_{-} H_0^{(2)}\left( \frac{1}{2} \tilde{k}_B \tilde{a}_B^2 \right) \right|^2,
		\end{align}
		where $\tilde{B}_{\pm} = \sqrt{a_1 H_1} B_{\pm}$, and
		\begin{align}
			\tilde{P}_B = \frac{3 P}{16 \pi^2 H_1^2}.
		\end{align}
	
	At $\eta > \eta_2$, the universe shifts into the slow-roll stage, and the solution is given by \eqref{eq:SolutionC}. To match the boundary between the eras, we fix the coefficients $C_{+}$ and $C_{-}$ by the continuity of $\mathcal{R}$ and $\mathcal{R}'$. One finds
		\begin{align}
			\label{eq:CoefficientCp} &\tilde{C}_{+} = \frac{e^{i \tilde{k}_C}}{2 \tilde{k}_C} \left\{ \left[ \tilde{k}_C H_{0, C}^{(1)} - ( 1 - i \tilde{k}_C ) H_{1, C}^{(1)} \right] \tilde{B}_{+} \right. \notag \\
			&\quad\quad\quad\quad\quad \left. + \left[ \tilde{k}_C H_{0, C}^{(2)} - ( 1 - i \tilde{k}_C ) H_{1, C}^{(2)} \right] \tilde{B}_{-} \right\}, \\
			\label{eq:CoefficientCm} &\tilde{C}_{-} = \frac{e^{-i \tilde{k}_C}}{2 \tilde{k}_C} \left\{ \left[ \tilde{k}_C H_{0, C}^{(1)} - ( 1 + i \tilde{k}_C ) H_{1, C}^{(1)} \right] \tilde{B}_{+} \right. \notag \\
			&\quad\quad\quad\quad\quad \left. + \left[ \tilde{k}_C H_{0, C}^{(2)} - ( 1 + i \tilde{k}_C ) H_{1, C}^{(2)} \right] \tilde{B}_{-} \right\},
		\end{align}
		where $H_{0, C}^{(1)}$ is the shorthand of $H_0^{(1)}( \tilde{k}_C /2 )$ and so on. The power spectrum in era C is given by \eqref{eq:PowerC} with $\tilde{C}_{+}$ and $\tilde{C}_{-}$ substituted by \eqref{eq:CoefficientCp} and \eqref{eq:CoefficientCm}, respectively.
	
\begin{figure}

\centering

\includegraphics[width = 8 cm]{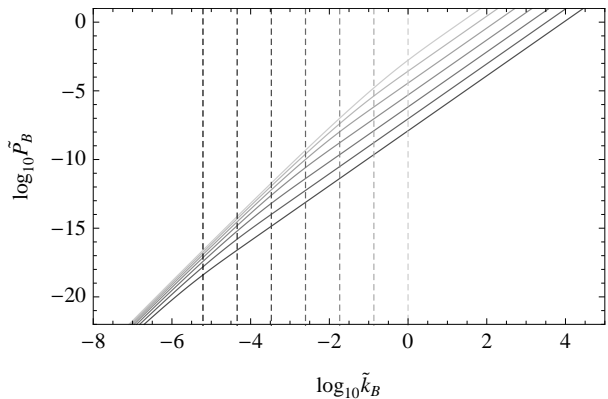}

\caption{Time evolution of the power spectrum in era B in the case of two-stage evolution (era B and C). The solid curves are the spectra of $\mathcal{R}$ from early time to late time (from light to dark). The vertical dashed lines denote the comoving horizon size at the corresponding instants (also from light to dark).}

\label{fig:BCinB70}

\end{figure}

	If the universe is in the adiabatic vacuum in era B, only the second term in \eqref{eq:SolutionB} presents in the mode function,
		\begin{align}
			u_B = \frac{1}{8 \pi} \tilde{a}_B H_0^{(2)}\left( \frac{1}{2} \tilde{k}_B \tilde{a}_B^2 \right),
		\end{align}
		where the normalization is chosen to recover the small-scale limit \eqref{eq:MinkowskiLimit}. In the kinetic era, the comoving horizon size increases with time, moving toward the left to the large scale (FIG.~\ref{fig:BCinB70}). The power drops after the mode enters the horizon, as in the slow-roll era, but it also decreases at the super-horizon scales before the horizon entry. This is actually a salient feature of the adiabatic vacuum in the decelerating era with $-1/3 < w < 1$.


\begin{figure}

\centering

\includegraphics[width = 8 cm]{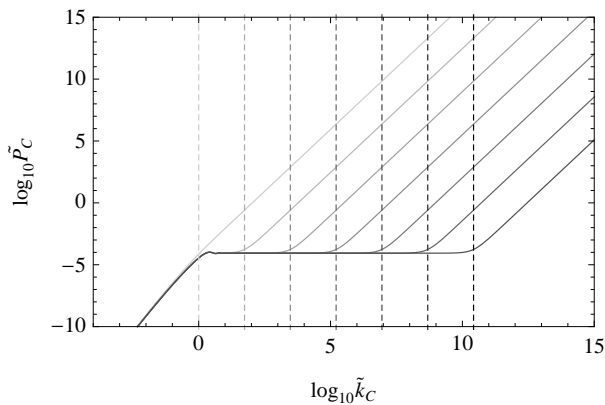}

\caption{Time evolution of the power spectrum in era C in the case of two-stage evolution (era B and C). The solid curves are the spectra of $\mathcal{R}$ from early time to late time (from light to dark). The vertical dashed lines denote the comoving horizon size at the corresponding instants (also from light to dark).}

\label{fig:BCinC70}

\end{figure}

	The evolution in the following slow-roll era (era C) demonstrates how the imprint of the initial vacuum is left at the large scales of the power spectrum at the end of inflation (modeled by era C). In the slow-roll era, the comoving horizon size decreases, going toward the right, and the modes exit the horizon (FIG.~\ref{fig:BCinC70}). The super-horizon modes have two different types of history. The ones with longest wavelengths have not entered the horizon yet at the end of era B, and stay outside the horizon in era C. These modes preserve the blue-tilted spectrum, and account for the power suppression induced by the kinetic era (cf.~\cite{Contaldi2003}). The other modes with shorter wavelengths enter the horizon in era B, and exit the horizon in era C. They are scale-invariant outside the horizon, as in the case of the single slow-roll scenario. This is because the adiabatic vacua approach the same short-wavelength limit \eqref{eq:MinkowskiLimit} in either the kinetic era or the slow-roll era. Therefore, although the spectrum has a scale-invariant segment due to the slow-roll era, the largest scales of the spectrum reveal the initial vacuum stemming from the kinetic era.


\begin{figure}

\centering

\includegraphics[width = 8.5 cm]{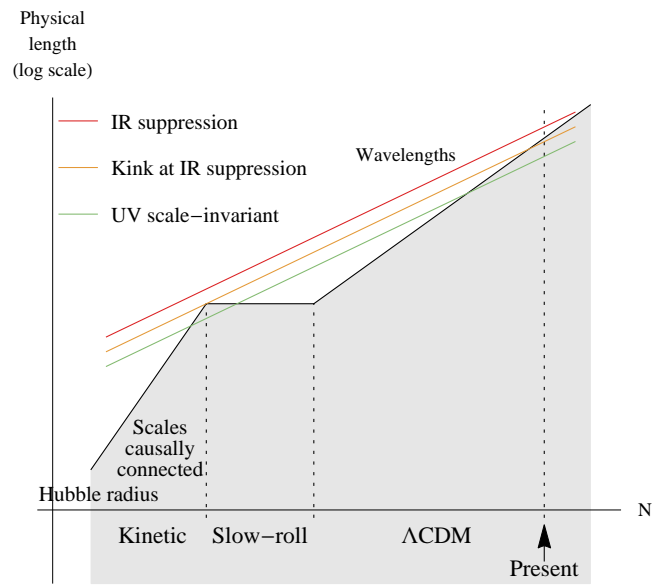}

\caption{Illustration of the evolution of the physical wavelengths of the modes and the Hubble radius with respect to the number of e-fold, $N$. The three parallel straight lines denote the modes with three different wavelengths, long to short from top to bottom (color online). The corresponding features they generate in the power spectrum, FIG.~\ref{fig:BCinC70}, are labeled in the legends (top to bottom corresponding to long to short wavelengths). The black piecewise-connected lines denote the Hubble radius evolving from the kinetic era to the slow-roll era, and finally into the $\Lambda$CDM era. The shaded region denotes the scales within which are causally connected.}

\label{fig:HorizonExitBC20}

\end{figure}

	The results we obtain crucially depend on the assumption about the initial vacuum of the universe. The prediction of power suppression is challenged by the fact that it originates from the blue-tilted super-horizon initial spectrum in the pre-inflation decelerating era, in which the super-horizon modes have not been in causal contact throughout the history. We demonstrate this point by an illustration showing the evolution of the Fourier wavelengths and the Hubble horizon (FIG.~\ref{fig:HorizonExitBC20}). In the decelerating universe, such as the initial kinetic era, the Hubble horizon grows faster than the Fourier wavelengths do. At the end of the initial decelerating era and the beginning of the accelerating inflation, the modes that are about to enter the horizon---those who are the origin of the suppressed large-scale modes today---will soon be expanded and kept outside the horizon by inflation. If the universe starts with the decelerating era before inflation, these mode are then insulated from any sub-horizon dynamics throughout the history. Therefore, the spectrum of the perturbations beyond the horizon size at the end of the decelerating era is not the consequence of causal physics, and the common approach of deducing the initial conditions through requiring the spectrum recover the Minkowski limit at the sub-horizon scale is therefore \emph{a posteriori}. This is of the same footing as the ``horizon problem'' the big bang cosmology faced before the picture of inflation was introduced.


	\subsection{Slow-roll---kinetic---slow-roll}
	
	\label{subsec:ABC}
	
		With the additional slow-roll era (era A) prepending the kinetic era (era B), we simply apply solution \eqref{eq:SolutionC} at $\eta < \eta_1$,
		\begin{align}
			\label{eq:SolutionA} u_A = \; &A_{+} \left( 1 - i \frac{\tilde{a}_A}{\tilde{k}_A} \right) e^{-i \tilde{k}_A / \tilde{a}_A} \notag \\
			&+ A_{-} \left( 1 + i \frac{\tilde{a}_A}{\tilde{k}_A} \right) e^{i \tilde{k}_A / \tilde{a}_A},
		\end{align}
		where $\tilde{k}_A = k / a_1 H_1$ and $\tilde{a}_A = a / a_1 = [ 1 - a_1 H_1 ( \eta - \eta_1 ) ]^{-1}$. The mode function of the adiabatic vacuum in era A is
		\begin{align}
			\label{eq:EraAVacuum} u_A = \frac{1}{(2 \pi)^{3/2} \sqrt{2 k}} \left( 1 + i \frac{\tilde{a}_A}{\tilde{k}_A} \right) e^{i \tilde{k}_A / \tilde{a}_A}.
		\end{align}
		Matching the boundary between era A and B, one finds the adiabatic vacuum in era A excites both modes of \eqref{eq:SolutionB} in era B with coefficients
		\begin{align}
			\tilde{B}_{+} = &-\frac{e^{i \tilde{k}_B}}{32 \sqrt{\pi \tilde{k}_B}} \left[ \tilde{k}_B H_{0, B}^{(2)} - ( 1 - i \tilde{k}_B ) H_{1, B}^{(2)} \right], \\
			\tilde{B}_{-} = &\frac{e^{i \tilde{k}_B}}{32 \sqrt{\pi \tilde{k}_B}} \left[ \tilde{k}_B H_{0, B}^{(1)} - ( 1 - i \tilde{k}_B ) H_{1, B}^{(1)} \right].
		\end{align}
		These results can be fed into \eqref{eq:CoefficientCp} and \eqref{eq:CoefficientCm}, identifying $H_1 = H_A$ as the Hubble constant in era A, and obtain the initial coefficients $\tilde{C}_{\pm}$ in era C. The power spectrum in era C is again given by \eqref{eq:PowerC} with the $\tilde{C}_{\pm}$ found.
	
\begin{figure}

\centering

\includegraphics[width = 8 cm]{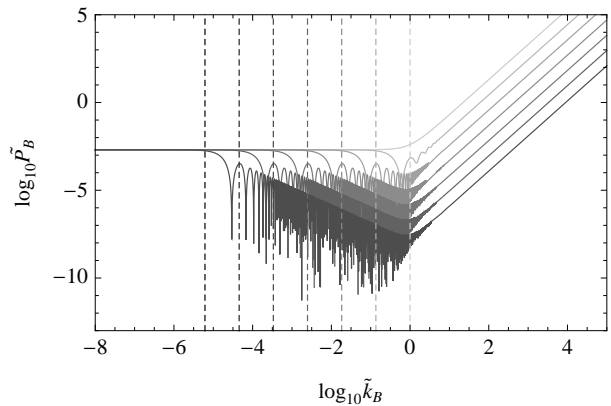}

\caption{Time evolution of the power spectrum in era B in the case of three-stage evolution (era A, B, and C). The solid curves are the spectra of $\mathcal{R}$ from early time to late time (from light to dark). The vertical dashed lines denote the comoving horizon size at the corresponding instants (also from light to dark).}

\label{fig:ABCinB70}

\end{figure}

	With the initial vacuum in the slow-roll era (era A), in which the spectrum evolves in the same way as it does in FIG.~\ref{fig:CinC70}, the super-horizon spectrum in era B is scale-invariant (FIG.~\ref{fig:ABCinB70}), different from the blue-tilted spectrum of the adiabatic vacuum in era B (FIG.~\ref{fig:BCinB70}). Moreover, after the modes enter the horizon, the power decrease and the spectrum becomes red-tilted (FIG.~\ref{fig:ABCinB70}), opposite to the blue-tilted spectrum in era B without era A (FIG.~\ref{fig:BCinB70}).


\begin{figure}

\centering

\includegraphics[width = 8 cm]{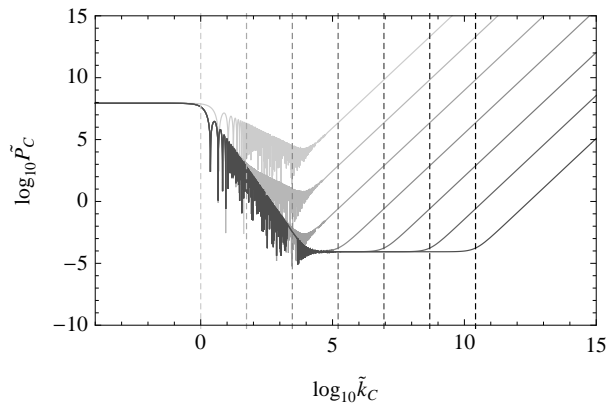}

\caption{Time evolution of the power spectrum in era C in the case of three-stage evolution (era A, B, and C). The solid curves are the spectra of $\mathcal{R}$ from early time to late time (from light to dark). The vertical dashed lines denote the comoving horizon size at the corresponding instants (also from light to dark).}

\label{fig:ABCinC70}

\end{figure}

	In era C, there are three different types of history inherited by the super-horizon modes (FIG.~\ref{fig:ABCinC70}). The modes with the longest wavelengths do not enter the horizon in era B and are kept outside of the horizon in era C. They preserve the scale-invariant spectrum as the proof of the existence of the initial vacuum in a slow-roll era (era A). For the modes with shorter wavelengths that enter the horizon in era B and exit the horizon in era C, the red-tilted shape of the spectrum persists, denting more as the power decreases inside the horizon. The super-horizon modes with shortest wavelengths exit the horizon for the first time in era C. These modes behave like the ones in era A, leaving the power scale-invariant as they exit the horizon. Note the magnitude of the scale-invariant spectrum generated in era C is lower than that generated in era A, since according to \eqref{eq:SlowRollSpectrum} the mode exiting the horizon from the adiabatic vacuum acquires the power that is proportional to the Hubble expansion rate, which is lower in era C.


\begin{figure}

\centering

\includegraphics[width = 8.5 cm]{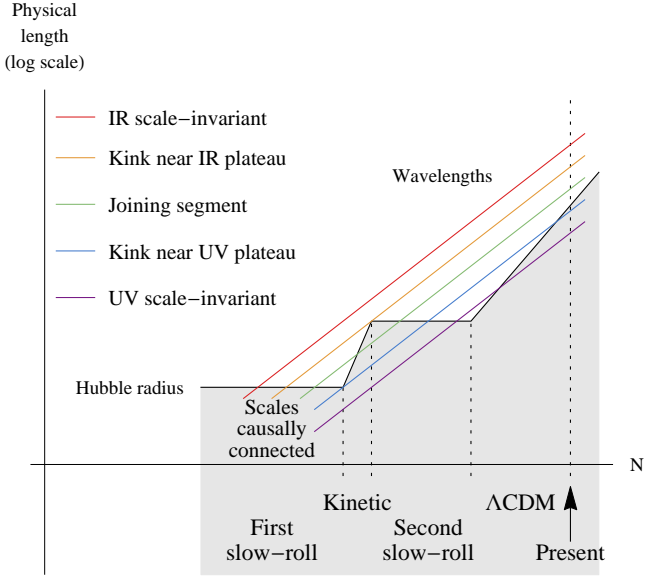}

\caption{Illustration of the evolution of the physical wavelengths of the modes and the Hubble radius with respect to the number of e-fold, $N$. The five parallel straight lines denote the modes with different wavelengths, long to short from top to bottom (color online). The corresponding features they generate in the power spectrum, FIG.~\ref{fig:ABCinC70}, are labeled in the legends (top to bottom corresponding to long to short wavelengths). The black piecewise-connected lines denote the Hubble radius evolving from the first slow-roll era to the kinetic era, then to the second slow-roll era, finally into the $\Lambda$CDM era. The shaded region denotes the scales within which are causally connected.}

\label{fig:HorizonExitABC40}

\end{figure}

	With the accelerating era A before the kinetic era B, all the modes are initially sub-horizon at the early times and are causal connected (FIG.~\ref{fig:HorizonExitABC40}). However, although this scenario is free from the acausal initial conditions, the intermediate kinetic era no longer leads to power suppression at the large scales. The accelerating era preceding the kinetic era changes the initial conditions, generating the the scale-invariant spectrum at the large scales. The power spectrum now has two scale-invariant segments: one generated in era A with larger power at the larger scales, and the other generated in era C with lower power at the smaller scales. The intermediate kinetic era in this case generates the spectrum that connects the two scale-invariant segments (FIG.~\ref{fig:ABCinC70}).


	We obtain three different large-scale behaviors from three different initial vacua and intermediate evolutions. Matching them with the inflation scenarios, we have the following interpretations. In the first case, with a single era C, the super-horizon spectrum is scale-invariant. This corresponds to the picture of inflation with large e-folding numbers (much larger than 60 e-folds). In the second case, the spectrum is suppressed at the large scales if the universe is in the adiabatic vacuum of era B before the onset of era C. This is the scenario of having a fast-roll era before the ``just enough'' inflation (with about 50--60 e-folds). In the third case, if before the fast-roll era (era B) the universe is in another slow-roll era (era A) at the early times, the spectrum is enhanced at the large scales. This corresponds to many supersymmetry or string motivated models that manifest the fast-roll era as a transient between two slow-roll eras \cite{Jain2009, Dudas2012, Yamauchi2011}.

	
	\subsection{Super-inflation---slow-roll}
	
	\label{subsec:SC}
	
		Consider the case when $\eta < \eta_2$ the universe is in the super-inflation era (era S), which is modeled by the Lagrangian \eqref{eq:PhantomLagrangian}. With $w < -1$, the mode function is
			\begin{align}
				\label{eq:modeS} u_S = &S_+ M_{0, \mu}\left( \frac{2 i \tilde{k}_S}{\alpha} \tilde{a}_S^{\alpha} \right) \notag \\
				&+ S_- W_{0, \mu}\left( \frac{2 i \tilde{k}_S}{\alpha} \tilde{a}_S^{\alpha} \right),
			\end{align}
		where $\tilde{k}_S = k / a_1 H_1$, $\tilde{a}_S = a / a_1 = [ 1 + \alpha a_1 H_1 ( \eta - \eta_1 ) ]^{1/\alpha}$, and $\alpha$ and $\mu$ are given by \eqref{eq:alpha} and \eqref{eq:mu}, respectively. Here $\eta_1 < \eta_2$ also denotes the reference time for era S. The normalized power spectrum is
			\begin{align}
				\label{eq:PowerS} \tilde{P}_S = &\frac{\tilde{k}_S^3}{\tilde{a}_S^2} \left| \tilde{S}_+ M_{0, \mu} \left( \frac{2 i \tilde{k}_S}{\alpha} \tilde{a}_S^{\alpha} \right) \right. \notag \\
				&\quad\quad \left. + \tilde{S}_- W_{0, \mu}\left( \frac{2 i \tilde{k}_S}{\alpha} \tilde{a}_S^{\alpha} \right) \right|^2,
			\end{align}
			where $\tilde{S}_{\pm} = \sqrt{a_1 H_1} S_{\pm}$ and
			\begin{align}
				\label{eq:PowerSDef} \tilde{P}_S = \frac{-(1+\alpha) P_S}{16 \pi^2 H_1^2}.
			\end{align}
		
		The universe goes into the slow-roll era (era C) at $\eta_2$, with mode function given by \eqref{eq:SolutionC}. Using the same matching conditions, the coefficients $\tilde{C}_{\pm}$ are found to be
		\begin{align}
			\label{eq:cp} \tilde{C}_+ = &\frac{\sqrt{\epsilon} e^{-\frac{1}{2} \alpha N_S} e^{i \tilde{k}_C}}{4 \sqrt{-(1+\alpha)} \tilde{k}_C^2} \notag \\
			&\times \left\{ 2 \left[ 2 \tilde{k}_C^2 + 2 i \tilde{k}_C - 1 \right] \tilde{S}_+ M_{0, \mu} \right. \notag \\
			&\quad + 2 \left[ 2 \tilde{k}_C^2 + 2 i \tilde{k}_C - 1 \right] \tilde{S}_- W_{0, \mu} \notag \\
			&\quad + \alpha ( 1 - i \tilde{k}_C ) ( 2 \mu + 1 ) \tilde{S}_+ M_{1, \mu} \notag \\
			&\quad \left. - 2 \alpha ( 1 - i \tilde{k}_C ) \tilde{S}_- W_{1, \mu} \right\}, \\
			\label{eq:cm} \tilde{C}_- = &\frac{\sqrt{\epsilon} e^{-\frac{1}{2} \alpha N_S} e^{-i \tilde{k}_C}}{4 \sqrt{-(1+\alpha)} \tilde{k}_C^2} \notag \\
			&\times \left\{ -2 \tilde{S}_+ M_{0, \mu} \right. \notag \\
			&\quad -2 \tilde{S}_- W_{0, \mu} \notag \\
			&\quad + \alpha ( 2 \mu + 1 ) ( 1 + i \tilde{k}_C ) \tilde{S}_+ M_{1, \mu} \notag \\
			&\quad \left. - 2 \alpha ( 1 + i \tilde{k}_C ) \tilde{S}_- W_{1, \mu} \right\},
		\end{align}
			with all Whittaker functions evaluated at $2 i \tilde{k}_C / \alpha$. The slow-roll parameter in era C is still given by $\epsilon$, and $N_S$ is the number of e-folds from $\eta_1$ to $\eta_2$.
		
\begin{figure}

\centering

\includegraphics[width = 8 cm]{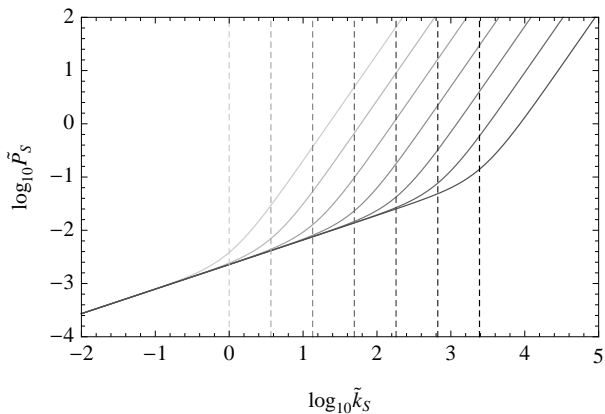}

\caption{Time evolution of the power spectrum in era S in the case of two-stage evolution (era S and C), with $w = -1.2$. The solid curves are the spectra of $\mathcal{R}$ from early time to late time (from light to dark). The vertical dashed lines denote the comoving horizon size at the corresponding instants (also from light to dark).}

\label{fig:SCinS22}

\end{figure}

		The adiabatic vacuum \eqref{eq:InitialCondition} in era S corresponds to the coefficients
			\begin{align}
				\tilde{S}_+ = &0, \\
				\tilde{S}_- = &\frac{\exp\left( \frac{i \tilde{k}_C}{\alpha} e^{-\alpha N_S} \right)}{(2\pi)^{3/2} \sqrt{2 \tilde{k}_C} e^{-\alpha N_S / 2}},
			\end{align}
			where we have used the relation $\tilde{k}_S = \tilde{k}_C e^{-\alpha N_S}$. The power spectrum in era S is plotted in FIG.~\ref{fig:SCinS22}, taking $w = -1.2$ as an example. Similar to the case of the kinetic era (era B), the super-horizon spectrum is blue-tilted, but in the super-inflation era the comoving Hubble radius decreases, and the super-horizon spectrum remains constant in time.
		
\begin{figure}

\centering

\includegraphics[width = 8 cm]{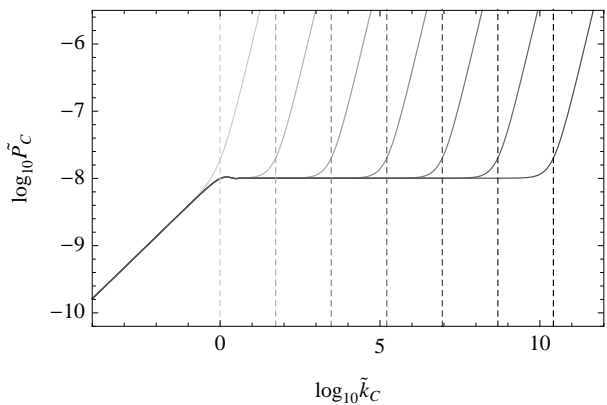}

\caption{Time evolution of the power spectrum in era C in the case of two-stage evolution (era S and C), with $w = -1.2$, $N_S = 6$, and $\epsilon = 0.1$. The solid curves are the spectra of $\mathcal{R}$ from early time to late time (from light to dark). The vertical dashed lines denote the comoving horizon size at the corresponding instants (also from light to dark).}

\label{fig:SCinC260}

\end{figure}

		As the universe enters the slow-roll era (era C), the super-horizon modes have two different types of history (FIG.~\ref{fig:SCinC260}). The modes with longer wavelengths exit the horizon in era S, retaining the blue-tilted super-horizon spectrum and staying constant in era C. The modes with shorter wavelengths exit the horizon in era C, acquiring the scale-invariant spectrum at horizon crossing.


\section{Two-field cascade inflation}

	\label{sec:TwoField}
	
	After establishing the understanding of the spectrum evolution in multi-stage inflation using the \emph{ad hoc} single-field analysis, in this section we calculate the spectrum generated by a two-stage inflation model from the given potential. The purposes are to show that the evolution pattern we obtain in the single-field analysis is also reflected in the two-field dynamics, and to check whether the large-scale power is suppressed due to the coupling between the two fields.
	
	We investigate a simple two-field cascade inflation, which is driven by a heavy scalar field $\psi$ and a light scalar field $\phi$ with the action
		\begin{align}
			\label{eq:action} S = \int d^4 x \sqrt{-g} \left[ -\frac{1}{2} \partial_{\alpha} \phi \partial^{\alpha} \phi -\frac{1}{2} \partial_{\alpha} \psi \partial^{\alpha} \psi - V(\phi, \psi) \right].
		\end{align}
		In large field inflation, the field operates at the super-Plankian scale when the coefficient of the kinetic term is normalized to $1/2$. If the field value is of the order of the Planck mass, $M_P = 1 / \sqrt{G}$, the energy scale of inflation is determined by the mass of the field. The typical evolution can therefore be divided into four stages. The first stage is the inflationary era driven by the heavy field, with fields starting far away from the potential minimum and rolling down alone the hillside of the potential. At the second stage the heavy field falls into the potential minimum and oscillates with damping amplitude. As the energy drops to the scale of the light field mass, the universe enters the third stage in which the light field initiates the other inflation era. Finally at the fourth stage the light field decays, ending the inflation, and the standard $\Lambda$CDM evolution begins.
	
	Consider the cascade inflation realized by the potential
		\begin{align}
			V(\phi, \psi) = \frac{\lambda'}{2} \phi^2 \psi^2 + \frac{1}{2} m^2 \phi^2.
		\end{align}
		The heavy-field inflation is driven by the coupling term $\lambda' \phi^2 \psi^2 / 2$, and the light-field inflation is driven by the mass term $m^2 \phi^2 / 2$. At the stage of heavy-field inflation, the fields follow the attractor solutions, which can be obtained through expressing the equations of motion in terms of the number of e-folds, $N \equiv \ln ( a / a_i )$, where $a_i$ is some initial scale factor. The Friedmann equation is
		\begin{align}
			\label{eq:Friedmann} H^2 = \frac{\frac{8 \pi}{3} V}{1 - \frac{4 \pi}{3} \left[ \left( \frac{d \phi}{d N} \right)^2 + \left( \frac{d \psi}{d N} \right)^2 \right]},
		\end{align}
		where $H = \dot{a} / a$. The dots denote the derivatives with respect to $t$. The equations of motion can be casted into the form
		\begin{align}
			\label{eq:phiEOMinN2} &\frac{d^2 \phi}{d N^2} - 4 \pi \left[ \left( \frac{d \phi}{d N} \right)^2 + \left( \frac{d \psi}{d N} \right)^2 - \frac{3}{4 \pi} \right] \notag \\
			&\quad\quad\quad\quad \times \left( \frac{d \phi}{d N} + \frac{1}{8 \pi V} \frac{\partial V}{\partial \phi} \right) = 0, \\
			\label{eq:psiEOMinN2} &\frac{d^2 \psi}{d N^2} - 4 \pi \left[ \left( \frac{d \phi}{d N} \right)^2 + \left( \frac{d \psi}{d N} \right)^2 - \frac{3}{4 \pi} \right] \notag \\
			&\quad\quad\quad\quad \times \left( \frac{d \psi}{d N} + \frac{1}{8 \pi V} \frac{\partial V}{\partial \psi} \right) = 0.
		\end{align}
		It can be shown that the attractor solutions satisfy
		\begin{align}
			\label{eq:phiAttractor} \frac{d \phi}{d N} = &-\frac{1}{8 \pi V} \frac{\partial V}{\partial \phi}, \\
			\label{eq:psiAttractor} \frac{d \psi}{d N} = &-\frac{1}{8 \pi V} \frac{\partial V}{\partial \psi},
		\end{align}
		provided $\left( d \phi / d N \right)^2 + \left( d \psi / d N \right)^2 - 3 / 4 \pi < 0$ as constrained by \eqref{eq:Friedmann}. With potential dominated by $\lambda' \phi^2 \psi^2 / 2$, the attractor solutions are
		\begin{align}
			\label{eq:phiStage1Case1} \phi(N) &= \sqrt{ \phi_i^2 - \frac{1}{2 \pi} ( N - N_i )}, \\
			\label{eq:psiStage1Case1} \psi(N) &= \sqrt{ \psi_i^2 - \frac{1}{2 \pi} ( N - N_i )},
		\end{align}
		where the subscripts $i$ denote the initial values.
		
		At the second stage, $\psi$ exits the slow-roll regime and oscillates at the minimum of the potential with its amplitude damped with time. The potential is still dominated by the coupling term before the next inflation begins. During the oscillatory stage, $\phi$ remains slow roll, while $\psi$ acquires a large kinetic energy that is of the same order of the potential energy, $\dot{\psi}^2 \sim \lambda' \phi^2 \psi^2 \sim H^2$, and the energy density evolves effectively according to $w = 0$ (zero pressure). Ignoring the kinetic energy of $\phi$ in the Friedmann equation, we have
		\begin{align}
			\label{eq:psiFriedmann} \left( \frac{d \psi}{d N} \right)^2 + \frac{\lambda' \phi^2}{H^2} \psi^2 = \frac{3}{4 \pi}.
		\end{align}
		With $A \equiv H / \sqrt{\lambda'} \phi$, we parametrize $\psi$ and $d \psi / d N$ by the amplitude $A$ and the phase $\theta$,
		\begin{align}
			\label{eq:psiPrimeDef} \frac{d \psi}{d N} = &\sqrt{\frac{3}{4 \pi}} \cos \theta, \\
			\label{eq:psiDef} \psi = &\sqrt{\frac{3}{4 \pi}} A \sin \theta.
		\end{align}
		Combining \eqref{eq:psiEOMinN2} and \eqref{eq:psiFriedmann}, we obtain a set of differential equations of $A$ and $\theta$
		\begin{align}
			\label{eq:DEA} \frac{d A}{d N} = &- A \cos^2 \theta \left( 3 + \frac{1}{\phi} \frac{d \phi}{d N} \right), \\
			\label{eq:DETheta} \frac{d \theta}{d N} = &\frac{1}{A} + \cos \theta \sin \theta \left( 3 + \frac{1}{\phi} \frac{d \phi}{d N} \right).
		\end{align}
		Among the two terms contributing to the frequency $d \theta / d N$, the first term $1 / A$ is of the order of $1 / \psi$, since $A = H / \sqrt{\lambda'} \phi \approx \sqrt{\lambda'} \phi \psi / \sqrt{\lambda'} \phi = \psi$. The second term is of order unity as $\phi$ slow rolls. After $\psi$ drops below the Planck mass, $1 / A$ dominates and $\theta$ oscillates rapidly, so we can approximate $\cos^2 \theta$ in \eqref{eq:DEA} by $1 / 2$. With $\phi$ given by the slow-roll solution \eqref{eq:phiStage1Case1}, $A$ and $\theta$ can then be integrated to yield
		\begin{align}
			A(N) &= A_i \left( 1 - \frac{N - N_i}{2 \pi \phi_i^2} \right)^{-1/4} e^{-\frac{3}{2} ( N - N_i )}, \\
			\theta(N) &= \theta_i + \frac{N - N_i}{A_i} - \frac{( N - N_i )^2}{16 \pi \phi_i^2 A_i}.
		\end{align}
	
	As the field $\psi$ attenuates, gradually the energy density from the coupling term is taken over by the $m^2 \phi^2 / 2$ term. The universe enters the third stage, in which the other inflation driven by the light-field begins. This stage is effectively described by the single-field inflation, with the same attractor solution \eqref{eq:phiStage1Case1} for $\phi$.

	
	Here we consider the perturbation to the homogeneous background in the conformal Newtonian gauge. The field perturbations are
		\begin{align}
			\phi(\mathbf{x}, t) = \bar{\phi}(t) + \delta \phi(\mathbf{x}, t), \\
			\psi(\mathbf{x}, t) = \bar{\psi}(t) + \delta \psi(\mathbf{x}, t),
		\end{align}
		where $\bar{\phi}(t)$ and $\bar{\psi}(t)$ denote the background solutions. The equations of motion of the perturbations in the Fourier space are
		\begin{align}
			\label{eq:deltaPhiEOM} &\delta \ddot{\phi} + 3 H \delta \dot{\phi} + \left( \frac{k^2}{a^2} + \frac{\partial^2 V}{\partial \phi^2} \right) \delta \phi = \notag \\
			&\quad\quad\quad\quad\quad\quad -\frac{\partial^2 V}{\partial \phi \partial \psi} \delta \psi -2 \frac{\partial V}{\partial \phi} \Psi + 4 \dot{\bar{\phi}} \dot{\Psi}, \\
			\label{eq:deltaPsiEOM} &\delta \ddot{\psi} + 3 H \delta \dot{\psi} + \left( \frac{k^2}{a^2} + \frac{\partial^2 V}{\partial \psi^2} \right) \delta \psi = \notag \\
			&\quad\quad\quad\quad\quad\quad -\frac{\partial^2 V}{\partial \phi \partial \psi} \delta \phi -2 \frac{\partial V}{\partial \psi} \Psi + 4 \dot{\bar{\psi}} \dot{\Psi}, \\
			&\label{eq:PsiEOM} \dot{\Psi} + H \Psi = 4 \pi ( \dot{\bar{\phi}} \delta \phi + \dot{\bar{\psi}} \delta \psi ).
		\end{align}
		Note that the energy-momentum tensor of action \eqref{eq:action} has no velocity perturbations either anisotropic inertia, so the Einstein equation gives $\Phi = \Psi$. The curvature perturbation in two-field system is
		\begin{align}
			\mathcal{R} = -\Psi - H \frac{ \dot{\bar{\phi}} \delta \phi + \dot{\bar{\psi}} \delta \psi }{ \dot{\bar{\phi}}^2 + \dot{\bar{\psi}}^2 }.
		\end{align}
	
	The initial conditions for the sub-horizon modes are set in the era of the heavy-field inflation by the method of iteration. We first neglect the metric perturbations $\Psi$ and find the solutions to $\delta \phi$ and $\delta \psi$ at the short-wavelength limit. The initial conditions for the sub-horizon field perturbations are set as the normalized positive-frequency solutions. We then feed the initial $\delta \phi$ and $\delta \psi$ back into the Einstein equations and obtain the initial conditions of $\Psi$. At the end we perform the consistency check to see whether the initial $\Psi$ obtained are indeed much smaller than $\delta \phi$ and $\delta \psi$ at the short-wavelength limit.
	
	To solve the field perturbations, first note that the equations of motion \eqref{eq:deltaPhiEOM} and \eqref{eq:deltaPsiEOM} can be put into a simpler form by substitute $t$ by the conformal time $\eta$, and introducing the new variables $w = a \delta \phi$ and $q = a \delta \psi$. Neglecting the metric perturbation, the equations for $w$ and $q$ are
		\begin{align}
			w'' + \left( k^2 + \frac{\partial^2 V}{\partial \phi^2} - \frac{a''}{a} \right) w = &- \frac{\partial^2 V}{\partial \phi \partial \psi} a^2 q, \\
			q'' + \left( k^2 + \frac{\partial^2 V}{\partial \psi^2} - \frac{a''}{a} \right) q = &- \frac{\partial^2 V}{\partial \phi \partial \psi} a^2 w,
		\end{align}
		where the primes denote the derivatives with respect to the conformal time $\eta$. For $k \gg a H$ the two equations decouple and reduce to the equations of harmonic oscillators
		\begin{align}
			w'' + k^2 w &= 0, \\
			q'' + k^2 q &= 0.
		\end{align}
		After the quantization, the normalized positive-frequency mode functions are
		\begin{align}
			\label{eq:InitialDeltaPhi} \delta \phi (\eta) = &\frac{e^{-i k \eta}}{(2 \pi)^{3/2} \sqrt{2 k} \; a}, \\
			\label{eq:InitialDeltaPsi} \delta \psi (\eta) = &\frac{e^{-i k \eta}}{(2 \pi)^{3/2} \sqrt{2 k} \; a},
		\end{align}
		which are the initial conditions for the sub-horizon field perturbations. Feeding \eqref{eq:InitialDeltaPhi} and \eqref{eq:InitialDeltaPsi} into the Einstein equation
		\begin{align}
			\label{eq:PsiRelation} &\Psi = \frac{1}{\dot{\bar{\phi}}^2 + \dot{\bar{\psi}}^2 - \frac{k^2}{4 \pi a^2}} \left[ \dot{\bar{\phi}} \delta \dot{\phi} + \dot{\bar{\psi}} \delta \dot{\psi} \right. \notag \\
			&\quad\quad \left. + \left( 3 H \dot{\bar{\phi}} + \frac{\partial V}{\partial \phi} \right) \delta \phi + \left( 3 H \dot{\bar{\psi}} + \frac{\partial V}{\partial \psi} \right) \delta \psi \right],
		\end{align}
		one obtains the initial conditions for the sub-horizon metric perturbations.
	
	To justify that the metric perturbations are negligible when finding solutions to the field perturbations, in this paragraph we are going to show that $\Psi$ obtained by \eqref{eq:PsiRelation} is much smaller than $\delta \phi$ and $\delta \psi$ at the short-wavelength limit in the era of heavy-field inflation. The following discussion in this paragraph assumes $k \gg a H$. First note that from \eqref{eq:phiStage1Case1} and \eqref{eq:psiStage1Case1} one has $\dot{\bar{\phi}} \sim H / \bar{\phi}$ and $\dot{\bar{\psi}} \sim H / \bar{\psi}$. The metric perturbation \eqref{eq:PsiRelation} therefore goes like
		\begin{align}
			\label{eq:PsiApprox1} \Psi \sim &\frac{1}{\left( \frac{k}{a H} \right)^2} \left[ \frac{\delta \dot{\phi}}{H \bar{\phi}} + \frac{\delta \dot{\psi}}{H \bar{\psi}} \right. \notag \\
			&\left. + \left( \frac{1}{\bar{\phi}} + \frac{\lambda' \bar{\phi} \bar{\psi}^2}{H^2} \right) \delta \phi + \left( \frac{1}{\bar{\psi}} + \frac{\lambda' \bar{\phi}^2 \bar{\psi}}{H^2} \right) \delta \psi \right],
		\end{align}
		where we have omitted all the constant coefficients and used the fact that the potential is dominated by $\lambda' \phi^2 \psi^2 / 2$ during the heavy-field inflation. From solutions \eqref{eq:InitialDeltaPhi} and \eqref{eq:InitialDeltaPsi} one has
		\begin{align}
			\label{eq:deltaPhiDot} \delta \dot{\phi} = - H \delta \phi \left( 1 + \frac{i k}{a H} \right) \sim \frac{k}{a} \delta \phi
		\end{align}
		and similarly for $\delta \dot{\psi}$. The first two terms in the bracket of \eqref{eq:PsiApprox1} then go like $( k / a H ) \cdot ( \delta \phi / \bar{\phi} )$ and $( k / a H ) \cdot ( \delta \psi / \bar{\psi} )$. Using the Friedmann equation we know that $\lambda' \bar{\phi} \bar{\psi}^2 / H^2 \sim 1 / \bar{\phi}$ and $\lambda' \bar{\phi}^2 \bar{\psi} / H^2 \sim 1 / \bar{\psi}$, therefore the last two terms in the bracket of \eqref{eq:PsiApprox1} go like $\delta \phi / \bar{\phi}$ and $\delta \psi / \bar{\psi}$. Since during the heavy-field inflation $\bar{\phi}$ and $\bar{\psi}$ is of order $\mathcal{O}(1)$ in Planck units, one has
		\begin{align}
			\Psi \sim \frac{1}{\left( \frac{k}{a H} \right)} \left( \frac{\delta \phi}{\bar{\phi}} + \frac{\delta \psi}{\bar{\psi}} \right).
		\end{align}
		Therefore the initial $\Psi$ is indeed much smaller than the initial $\delta \phi$ and $\delta \psi$ at the limit of $k \gg a H$ in the era of heavy-field inflation.
	
	The CMB spectrum is found by three steps of numerical calculations. We first obtain the background dynamics by solving \eqref{eq:phiEOMinN2} and \eqref{eq:psiEOMinN2}. The reheating energy scale is assumed to be $7.19 \times 10^{-4} M_P \sim 10^{16} \textrm{GeV}$. The evolution of perturbations is then solved by integrating \eqref{eq:deltaPhiEOM}, \eqref{eq:deltaPsiEOM}, and \eqref{eq:PsiEOM}, with the initial conditions set by \eqref{eq:InitialDeltaPhi}, \eqref{eq:InitialDeltaPsi}, and \eqref{eq:PsiRelation}. The spectrum of the curvature perturbations is found by evolving each Fourier mode until it reaches the steady value after the horizon exit. The resulting spectrum is then fed into \texttt{CAMB} \cite{Lewis2000, Lewis2012}, which is modified to accept arbitrary initial spectrum represented by an interpolating function, to calculate the spectrum of the CMB temperature fluctuations.
	
	There are two parameters and four initial conditions in our model. The two parameters are the coupling constant, $\lambda'$, and the mass of the light field, $m$. The four initial conditions are the initial values and derivatives of the fields $\phi$ and $\psi$. We assume that the light field drives about the last 60 e-folds of inflation, so the mass $m$ is determined by the Hubble scale during inflation deduced by the observation. The initial derivatives of $\phi$ and $\psi$ are set to zero, leaving the system released from rest and evolving into the attractor solutions.
	
\begin{figure}

\center

\includegraphics[width = 8 cm]{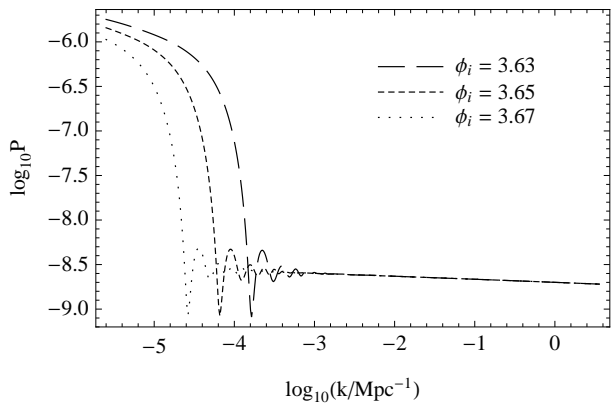}

\caption{The spectra of the curvature perturbations with $\phi_i = 3.63 M_P$, $3.65 M_P$, and $3.67 M_P$. The other parameters are held fixed as $\lambda' = 10^{-9}$, $\psi_i = 1.60 M_P$, and $m = 1.22 \times 10^{-6} M_P$.}

\label{fig:ChiSpectrum560}

\end{figure}

\begin{figure}

\center

\includegraphics[width = 8 cm]{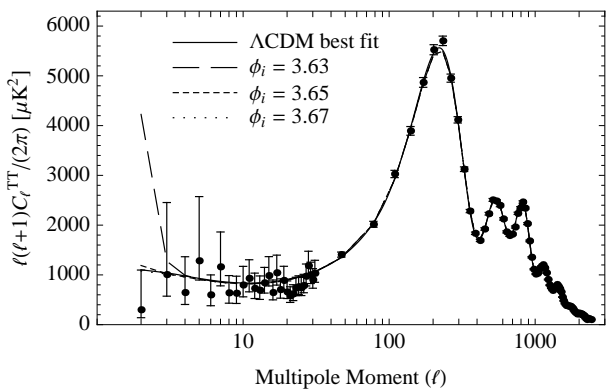}

\caption{The CMB temperature-temperature correlation (TT) spectra with $\phi_i = 3.63 M_P$, $3.65 M_P$, and $3.67 M_P$. The dots with error bars are the Planck 2013 data. The other parameters are held fixed as $\lambda' = 10^{-9}$, $\psi_i = 1.60 M_P$, and $m = 1.22 \times 10^{-6} M_P$.}

\label{fig:CMB553ScanPhi}

\end{figure}

	The initial value of the light field $\phi$ determines the behavior of the system in two ways. First, it determines the energy scale of the heavy-field inflation as well as that of the oscillatory period, since at the beginning of the oscillatory stage $\psi \sim M_P$ and $H \sim \sqrt{\lambda'} \phi$. Second, it determines the number of e-folds of the light-field inflation. With larger initial $\phi$, the light-field inflation begins at a higher energy scale and lasts longer. In this case, only those modes with larger wavelengths that are affected by the heavy-field inflation and the oscillatory stage (FIG.~\ref{fig:ChiSpectrum560}). Therefore, less deviations are manifested in the present-day CMB spectrum since those large modes have not entered the Hubble horizon today (FIG.~\ref{fig:CMB553ScanPhi}).


\begin{figure}

\center

\includegraphics[width = 8 cm]{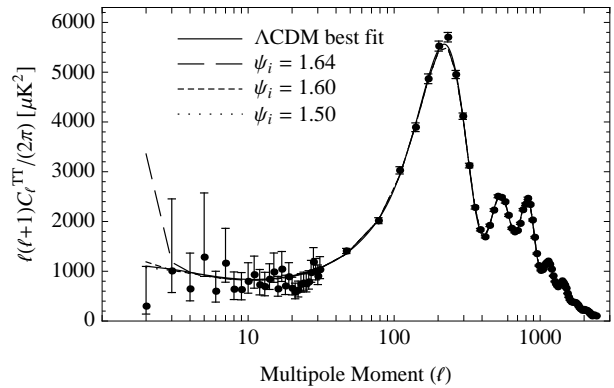}

\caption{The CMB TT spectra with $\psi_i = 1.50 M_P$, $1.60 M_P$, and $1.64 M_P$. The dots with error bars are the Planck 2013 data. The other parameters are held fixed as $\lambda' = 10^{-9}$, $\phi_i = 3.65 M_P$, and $m = 1.22 \times 10^{-6} M_P$.}

\label{fig:CMB562ScanPsi}

\end{figure}

	Raising the value of initial $\psi$ affects the spectrum by making the heavy-field inflation longer and, while holding the initial $\phi$ fixed, the light-field inflation shorter, without changing the duration of the transition era. The larger the initial $\psi$ is, more visible $k$ modes and $\ell$ modes are affected (FIG.~\ref{fig:CMB562ScanPsi}).


\begin{figure}

\center

\includegraphics[width = 8 cm]{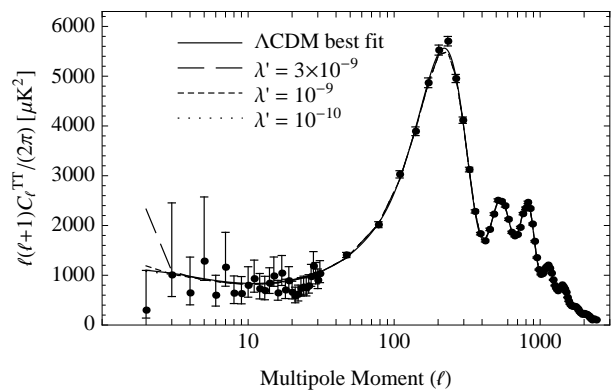}

\caption{The CMB TT spectra with $\lambda' = 10^{-10}$, $10^{-9}$, and $3 \times 10^{-9}$. The dots with error bars are the Planck 2013 data. The other parameters are held fixed as $\phi_i = 3.65 M_P$, $\psi_i = 1.60 M_P$, and $m = 1.22 \times 10^{-6} M_P$.}

\label{fig:CMB522ScanLambdaPrime}

\end{figure}

	The coupling $\lambda'$ determines the strength of the interaction between the two fields. With stronger interactions, it takes longer for the transient oscillation to settle, and therefore more modes with short wavelengths are affected while other conditions held fixed (FIG.~\ref{fig:CMB522ScanLambdaPrime}).


	We see from both the spectra of the curvature perturbations (FIG.~\ref{fig:ChiSpectrum560}) and CMB (FIG.~\ref{fig:CMB553ScanPhi}) that the large-scale spectrum is determined by the initial era with which the universe begins: The large-scale spectrum reflects the scale-invariant shape of the super-horizon spectrum in the initial slow-roll (heavy-field inflation) era. Although the oscillatory stage behaves like $w = 0$, the spectrum evolution is qualitatively similar to that shown in Subsec.~\ref{subsec:ABC}. A quantitative analysis of the spectrum evolution involving the zero-pressure ($w = 0$) era is given in the Appendix, showing that the \emph{ad hoc} single-field analysis does capture the pattern of the spectrum evolution and is in agreement with the numerical results.


\section{Conclusions}

	\label{sec:Conclusion}
	
	We show that if the universe begins in the super-inflation era ($w < -1$) or that with positive pressure ($w > 0$), the large-scale curvature perturbation spectrum is suppressed due to the blue-tilted super-horizon spectrum in the initial era. We first find the scaling relation of the super-horizon spectrum for a scalar field in the FLRW background with constant equation of state. At the large scales, the spectrum is blue-tilted for positive-pressure ($w > 0$) or super-inflation ($w < -1$) era, and red-tilted for the era with $-1 < w < 0$, except the singular case with $w = -1/3$. In the slow-roll ($w \simeq -1$) and zero-pressure ($w = 0$) background, the super-horizon spectrum is scale-invariant. We also point out that the conclusions are drawn from assuming the mode function approaches the Minkowski limit at small scales. Although being natural in the accelerating universe, this assumption becomes \emph{a posteriori} in the decelerating universe since the sub-horizon modes are evolved from the super-horizon modes, which are initially across causally disconnected regions before entering the horizon.
	
	By analyzing three scenarios:~a single slow-roll era, a slow-roll era preceded by a kinetic era, and two successive slow-roll eras connected by a kinetic era, we show the following two facts. First, the large-scale power suppression in the model with a single pre-inflation kinetic era stems from the blue-tilted super-horizon spectrum of the initial kinetic era. Second, the additional slow-roll era preceding the kinetic era changes the super-horizon initial spectrum, so the large-scale power is enhanced, rather than suppressed. These results show that the large-scale spectrum depends sensitively on the initial vacuum. In the universe beginning with the positive-pressure era, as we pointed out earlier, the super-horizon modes are initially across causally disconnected regions, and the well-motivated assumption on the initial state is still lacking. Some investigations about the effect of the different initial vacuum on the spectrum have been carried out in the literature \cite{Sriramkumar2005, Boyanovsky2006, Holman2008, Agullo2011}. We also explore the case that is free from the acausal issue: a super-inflation era preceding the slow-roll era, and show that the large-scale spectrum is suppressed due to the initially blue-tilted spectrum in the super-inflation era.
	
	We calculate the curvature perturbation and CMB spectra of a two-stage inflation model from the given two-field potential. We show that the large-scale power is enhanced due to the initial spectrum set in the first accelerating era, and the effect of the intermediate decelerating era on the spectrum is connecting the two plateaus generated in the two accelerating eras, which agrees with the picture obtained through the \emph{ad hoc} single-field analysis.


\begin{acknowledgements}

	We are grateful for the discussions with F.~Arroja, S.~Downes, C.~Gauthier, J.~Gu, K.~Izumi, L.~Labun, A.~Linde, A.~Mazumdar, T.~Qiu, M.~Sasaki, L.~Senatore, T.~Suyama, and D.~Yeom. Y.~L.~also want to thank KIPAC, SLAC National Accelerator Laboratory and SITP, Stanford University for the hospitality. P.~C.~and Y.~L.~are supported by Taiwan National Science Council under Project No.~NSC 97-2112-M-002-026-MY3 and by Taiwan National Center for Theoretical Sciences (NCTS). P.~C.~is in addition supported by U.S.~Department of Energy under Contract No.~DE-AC03-76SF00515. Y.~L.~is in addition supported by the Graduate Students Study Abroad Program No.~104-2917-I-002-008 sponsored by Taiwan Ministry of Science and Technology.

\end{acknowledgements}


\appendix*

\section{Spectrum evolution involving a zero-pressure era}
	
	Consider the universe consists of three successive eras: a first slow-roll era (era A), an intermediate zero-pressure ($w = 0$) era (era B), and a second slow-roll era (era C). Much parallel to the discussion in Subsec.~\ref{subsec:ABC}, we denote the quantities at the transitions from era A to B and from era B to C by subscripts 1 and 2, respectively.
	
	In era A, the mode function of the initial adiabatic vacuum is given by \eqref{eq:EraAVacuum}. In era B, we first keep the discussion general for $w > -1$. The mode function $u$ is given by \eqref{eq:WhittakerSolution},
			\begin{align}
				\label{eq:modeM} u_B = &B_+ M_{0, \mu}\left( \frac{2 i \tilde{k}_B}{\alpha} \tilde{a}_B^{\alpha} \right) \notag \\
				&+ B_- W_{0, \mu}\left( \frac{2 i \tilde{k}_B}{\alpha} \tilde{a}_B^{\alpha} \right),
			\end{align}
			where $\tilde{k}_B = k / a_1 H_1$ and $\tilde{a}_B = a / a_1 = [ 1 + \alpha a_1 H_1 ( \eta - \eta_1 ) ]^{1/\alpha}$. Parameters $\alpha$ and $\mu$ are given by \eqref{eq:alpha} and \eqref{eq:mu}, respectively. The normalized power spectrum is
			\begin{align}
				\label{eq:PowerM} \tilde{P}_B = &\frac{\tilde{k}_B^3}{\tilde{a}_B^2} \left| \tilde{B}_+ M_{0, \mu} \left( \frac{2 i \tilde{k}_B}{\alpha} \tilde{a}_B^{\alpha} \right) \right. \notag \\
				&\quad\quad \left. + \tilde{B}_- W_{0, \mu}\left( \frac{2 i \tilde{k}_B}{\alpha} \tilde{a}_B^{\alpha} \right) \right|^2,
			\end{align}
			where $\tilde{B}_{\pm} = \sqrt{a_1 H_1} B_{\pm}$ and
			\begin{align}
				\label{eq:PowerMDef} \tilde{P}_B = \frac{(1+\alpha) P_B}{16 \pi^2 H_1^2}.
			\end{align}
			In era C, the mode function is \eqref{eq:SolutionC}, and the normalized power spectrum is \eqref{eq:PowerC}. For simplicity, we assume that the slow-roll parameter $\epsilon$ in era A and C has the same value.
	
	By matching at the boundary $\eta = \eta_1$, we obtain
		\begin{align}
			\tilde{B}_+ = &\Delta_B \left[ (-2 i \tilde{k}_B^2 + 2 \tilde{k}_B + i) W_{0,\mu} \right. \notag \\
			&\quad\quad \left. + \alpha (\tilde{k}_B + i) W_{1, \mu} \right], \\
			\tilde{B}_- = &\Delta_B \left[ -(-2 i \tilde{k}_B^2 + 2 \tilde{k}_B + i) M_{0,\mu} \right. \notag \\
			&\quad\quad \left. + \frac{2 \mu + 1}{2} \alpha (\tilde{k}_B + i) M_{1, \mu} \right],
		\end{align}
		where
		\begin{align}
			\Delta_B = &\frac{\sqrt{\alpha + 1} e^{i \tilde{k}_B}}{2 (\pi \tilde{k}_B)^{3/2} \alpha \sqrt{\epsilon}} \notag \\
			&\times \frac{1}{(2 \mu + 1) M_{1,\mu} W_{0, \mu} + 2 M_{0,\mu} W_{1, \mu}},
		\end{align}
		and the Whittaker functions are evaluated at $2 i \tilde{k}_B / \alpha$. Matching at $\eta = \eta_2$ gives
		\begin{align}
			\label{eq:cp2} \tilde{C}_+ = &\frac{\sqrt{\epsilon} e^{-\frac{1}{2} \alpha N_B} e^{i \tilde{k}_C}}{4 \sqrt{1+\alpha} \; \tilde{k}_C^2} \notag \\
			&\times \left\{ 2 \left[ 2 \tilde{k}_C^2 + 2 i \tilde{k}_C - 1 \right] \tilde{B}_+ M_{0, \mu} \right. \notag \\
			&\quad + 2 \left[ 2 \tilde{k}_C^2 + 2 i \tilde{k}_C - 1 \right] \tilde{B}_- W_{0, \mu} \notag \\
			&\quad + \alpha ( 1 - i \tilde{k}_C ) ( 2 \mu + 1 ) \tilde{B}_+ M_{1, \mu} \notag \\
			&\quad \left. - 2 \alpha ( 1 - i \tilde{k}_C ) \tilde{B}_- W_{1, \mu} \right\}, \\
			\label{eq:cm2} \tilde{C}_- = &\frac{\sqrt{\epsilon} e^{-\frac{1}{2} \alpha N_B} e^{-i \tilde{k}_C}}{4 \sqrt{1+\alpha} \; \tilde{k}_C^2} \notag \\
			&\times \left\{ -2 \tilde{B}_+ M_{0, \mu} \right. \notag \\
			&\quad -2 \tilde{B}_- W_{0, \mu} \notag \\
			&\quad + \alpha ( 2 \mu + 1 ) ( 1 + i \tilde{k}_C ) \tilde{B}_+ M_{1, \mu} \notag \\
			&\quad \left. - 2 \alpha ( 1 + i \tilde{k}_C ) \tilde{B}_- W_{1, \mu} \right\},
		\end{align}
			with the Whittaker functions evaluated at $2 i \tilde{k}_C / \alpha$. One also has the relation $\tilde{k}_B = \tilde{k}_C e^{-\alpha N_B}$.
	
\begin{figure}

\centering

\includegraphics[width = 8 cm]{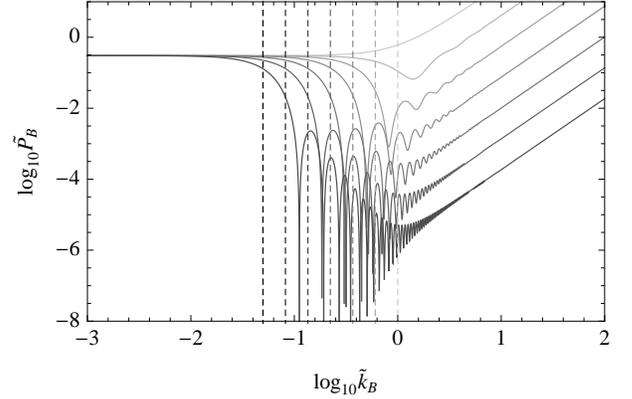}

\caption{Time evolution of the power spectrum in era B in the case of three-stage evolution (era A, B, and C), in which $w = 0$ in era B. The solid curves are the spectra of $\mathcal{R}$ from early time to late time (from light to dark). The vertical dashed lines denote the comoving horizon size at the corresponding instants (also from light to dark).}

\label{fig:AMCinM40}

\end{figure}

\begin{figure}

\centering

\includegraphics[width = 8 cm]{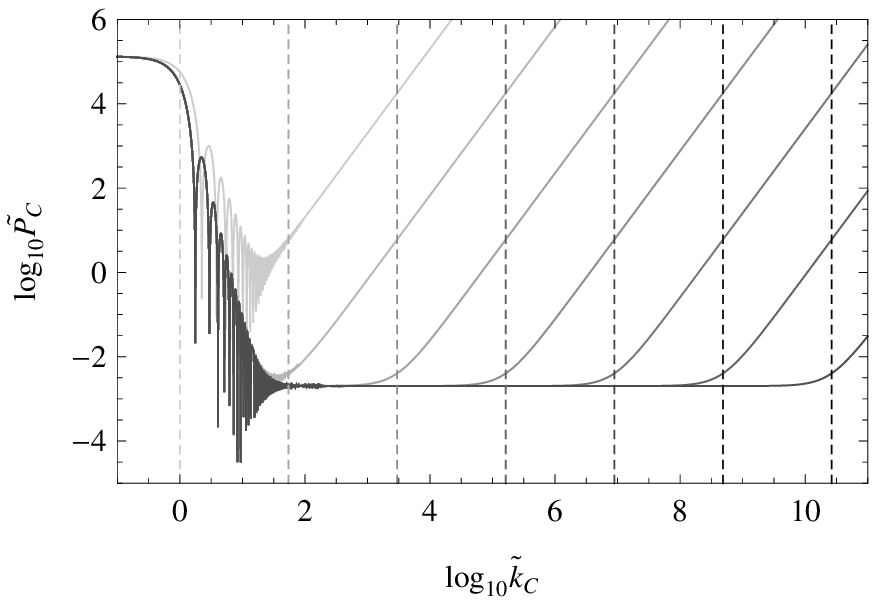}

\caption{Time evolution of the power spectrum in era C in the case of three-stage evolution (era A, B, and C), in which $w = 0$ in era B. The solid curves are the spectra of $\mathcal{R}$ from early time to late time (from light to dark). The vertical dashed lines denote the comoving horizon size at the corresponding instants (also from light to dark).}

\label{fig:AMCinC271}

\end{figure}

	Setting $w = 0$ in era B, we find the spectrum evolution as FIG.~\ref{fig:AMCinM40}. Similar to the case of the intermediate kinetic era in Subsec.~\ref{subsec:ABC}, the super-horizon modes inherits the scale-invariant spectrum in the initial slow-roll era (era A), and the modes that enter the horizon during the zero-pressure era (era B) have the red-tilted spectrum. Entering into the second slow-roll era (era C), as shown in FIG.~\ref{fig:AMCinC271}, the modes that enter the horizon in era B are expelled out of the horizon again, leading to the steep red-tilted spectrum connecting the two plateaus. The right plateau is formed by the modes that are sub-horizon during eras A and B and then exit the horizon in era C. Comparing to the spectrum of the two-field model (FIG.~\ref{fig:ChiSpectrum560}), which also has a zero-pressure era between two inflationary eras, we see that the \emph{ad hoc} single-field analysis agrees with the numerical result and largely captures the physics of the spectrum formation.


\bibliography{Cosmology}

\end{document}